\documentclass[11pt]{article}
\usepackage[margin=1in,letterpaper]{geometry}
\usepackage{algpseudocode}
\usepackage{graphicx}
\usepackage[bf,sf]{titlesec}
\usepackage{setspace}
\usepackage{titling}
\usepackage{natbib}
\usepackage{amsmath}
\usepackage{amssymb}
\usepackage[table,dvipsnames]{xcolor}
\usepackage{algorithm}
\usepackage{comment}
\usepackage[noEnd=True]{algpseudocodex}
\usepackage{multirow}
\usepackage{booktabs}
\usepackage{makecell}
\usepackage{subcaption}
\usepackage{fancybox}
\usepackage{fancyvrb}
\usepackage[colorlinks]{hyperref}
\usepackage{pdfx}
\usepackage{url}
\usepackage{adjustbox}
\usepackage{algorithmicx}
\usepackage{algpseudocode}
\usepackage{bm}
\usepackage{bbm}
\usepackage{pifont}
\usepackage{xspace}
\usepackage{wrapfig}
\usepackage{enumitem}
\usepackage{listings}
\usepackage{xurl}
\usepackage{fvextra}

\newcommand{\myparatight}[1]{\smallskip\noindent{\bf {#1}:}~}
\newcommand{\textbsf}[1]{\textsf{\textbf{#1}}}

\makeatletter
\renewenvironment{abstract}{%
    \if@twocolumn
      \section*{\abstractname}%
    \else %
      \begin{center}%
        {\sffamily \bfseries \abstractname\vspace{\z@}}%
      \end{center}%
      \quotation
    \fi}
    {\if@twocolumn\else\endquotation\fi}
\makeatother

\hypersetup{
  colorlinks   = true, %
  urlcolor     = RoyalBlue, %
  linkcolor    = RoyalBlue, %
  citecolor   = RoyalBlue %
}

\title{\textbsf{Evaluating Tool Cloning in Agentic-AI  Ecosystems}}
\author{
Taein Kim$^*$, David Jiang$^*$, Yuepeng Hu, Yuqi Jia, Neil Gong
\\
Duke University
\\
\texttt{\{taein.kim, david.jiang, yuepeng.hu, yuqi.jia, neil.gong\}@duke.edu}
\\
$^*$Equal contribution.
}
\date{}

\begin{document}

\maketitle

\begin{abstract}
Agent tools are becoming a core interface through which LLM agents access external data, services, and execution environments. As these tools are distributed through public marketplaces, raw tool counts may substantially overstate ecosystem diversity if many repositories are cloned, lightly modified, or derived from shared templates. Such hidden duplication can contaminate benchmark splits, propagate vulnerable implementations, bias measurements of tool-use generalization, and raise provenance, attribution, and intellectual-property concerns. We present, to our knowledge, the first large-scale measurement study of tool cloning in agentic AI ecosystems. We curate a unified dataset from multiple public platforms, covering 7,508 Model Context Protocol (MCP) repositories with 87,564 extracted tools and 1,353 Skills repositories with 12,447 tools, for a total of 8,861 repositories and 100,011 tool entries. To measure implementation-level duplication, we build a repository-level auditing pipeline using complementary lexical and fuzzy-structural similarity metrics, and compute pairwise similarity across MCP-to-MCP, Skills-to-Skills, and MCP-to-Skills repository pairs. We further manually verify 100 sampled pairs per MCP and Skills ecosystem across similarity-score buckets to calibrate how often high similarity reflects true code cloning. Our analysis shows that cloning is not an isolated artifact: high-similarity regions appear across comparison settings, and 60\% of high-Jaccard candidates and 85\% of high-\texttt{ssdeep} candidates in the MCP ecosystem are manually verified as clones. These results indicate that tool cloning is a pervasive and severe source of hidden duplication in agent-tool ecosystems. They further suggest that agent-tool datasets and benchmarks should account for repository provenance and implementation similarity when measuring tool diversity or constructing evaluation splits.
\end{abstract}

\section{Introduction}
LLM agents are increasingly used to interact with external data, services, and execution environments. Unlike standalone language models that primarily generate text, LLM agents can invoke tools to query databases, access files, call APIs, operate over code repositories, and execute task-specific procedures~\citep{yao2023react,shinn2023reflexion,shen2023hugginggpt,schick2023toolformer,yao2022webshop,liuagentbench,jimenez2023swe}. To support such interactions, recent agent frameworks expose tools through model-facing interfaces, including tool names, natural-language descriptions, input schemas, and executable backends. Representative examples include the Model Context Protocol (MCP)~\citep{anthropic_mcp_2024}, which provides standardized access to external tools and resources, and Skills repositories~\citep{anthropic_skills_2025}, which package reusable instructions or procedures for agentic tasks.

The rapid growth of these tool platforms has made agent tools an important part of modern agentic-AI infrastructure. Public marketplaces allow developers to publish tools at scale, and the resulting repositories are increasingly used for tool discovery, agent evaluation, and ecosystem analysis. However, the number of listed tools does not necessarily reflect the number of independent implementations. Many tools may be wrappers around similar external services, instantiated from common server templates, or lightly modified from existing tools. As a result, an ecosystem may appear large while being partially composed of cloned or template-derived implementations.

This hidden duplication creates several risks. For evaluation, tool-use benchmarks often treat different tools or repositories as independent instances~\citep{tang2023toolalpaca,qin2023toolllm}. If repository-level cloning is common, random benchmark splits may place similar implementations across training and test sets. A model may then appear to generalize to unseen tools while encountering familiar implementation patterns, schemas, or execution logic. For security, audits that inspect tools independently may overlook the propagation of the same vulnerable scaffold or unsafe code pattern across many repositories. For ecosystem governance, copied tools also raise provenance, attribution, and license-compliance concerns, including potential intellectual-property infringement when repositories are redistributed without permission or proper attribution. Yet the severity of these risks remains an open empirical question: existing work has not systematically measured whether tool cloning in real-world agent-tool ecosystems is rare and isolated, or widespread enough to affect ecosystem diversity, benchmarking, and security auditing.

To answer this question, we conduct a large-scale measurement study of tool cloning in agent-tool ecosystems. We curate a unified dataset from multiple public platforms, covering 7,508 MCP  repositories with 87,564 extracted tools and 1,353 Skills repositories with 12,447 tools. We then build a repository-level auditing pipeline to measure implementation similarity using two complementary signals: token-level Jaccard similarity and fuzzy structural similarity via \texttt{ssdeep}. We compute pairwise similarity across MCP-to-MCP, Skills-to-Skills, and cross-domain MCP-to-Skills repository pairs. Because similarity scores alone do not establish cloning, we manually verify sampled pairs from different similarity-score buckets and use the verified clone rates to calibrate how often high similarity corresponds to true implementation reuse. This design allows us to measure not only whether high-similarity pairs exist, but also whether they represent true clones and how broadly such clone candidates appear across the ecosystem.

Our results show that code cloning is not an isolated artifact but a pervasive source of hidden duplication in agent-tool ecosystems. High-similarity regions appear across MCP-to-MCP, Skills-to-Skills, and cross-domain MCP-to-Skills comparisons, and manual verification confirms that these regions are substantially enriched for true clones. These findings indicate that raw repository and tool counts can overstate effective ecosystem diversity, and that provenance-aware analysis is necessary for reliable benchmark construction and security auditing.

The contributions of this work are summarized as follows:
\begin{itemize}
    \item We curate a large-scale dataset of agent-tool repositories and metadata from multiple public distribution platforms, covering MCP servers and Skills.
    \item We develop a repository-level auditing pipeline for measuring lexical and structural similarity across agent-tool implementations.
    \item We manually verify repository pairs across similarity-score buckets and show that high-similarity regions are enriched for true clones, demonstrating that tool cloning is a pervasive and severe source of hidden duplication in the ecosystem.
    \item We discuss the implications of tool cloning for benchmark construction, provenance-aware dataset splitting, security auditing, and license-compliance analysis of agent-tool ecosystems.
\end{itemize}
\section{Related Work}
\subsection{LLM Agents and Tool Ecosystems}
Recent work has extended large language models from standalone text generators into agentic systems that can invoke external tools, access external state, and interact with software environments~\citep{yao2023react,shinn2023reflexion,shen2023hugginggpt,schick2023toolformer,liuagentbench,jimenez2023swe}. In these systems, tools provide model-accessible interfaces to external functionality, such as web services, databases, code repositories, file systems, and execution environments. At runtime, an agent selects tools based on their names, descriptions, schemas, and the current user request, then uses tool responses to continue reasoning or complete the task~\citep{anthropic_mcp_2024, anthropic_skills_2025}. This design has become a central abstraction for building practical LLM agents, as it allows models to operate beyond their parametric knowledge and interact with dynamic external systems.

A growing body of work studies how well LLM agents can use tools. Existing benchmarks evaluate whether models can select appropriate tools, generate valid arguments, interpret tool outputs, and compose multiple tool calls across complex tasks~\citep{qin2023toolllm,tang2023toolalpaca,patil2024gorilla}. These studies primarily focus on the model side of tool use: whether an agent can plan over available tools and invoke them correctly. In contrast, the structure of the tool ecosystem itself has received comparatively less attention. Existing evaluations often treat tools or tool repositories as independent functional units, without explicitly accounting for whether their implementations are genuinely distinct or derived from shared templates.

This assumption becomes increasingly important as agent tools are distributed through public repositories, marketplaces, and package-like ecosystems. Developers may adapt existing tools, instantiate common scaffolds, or wrap similar APIs with only minor implementation changes. As a result, the apparent scale of an agent-tool ecosystem may overestimate its true implementation diversity. Such duplication has implications for both evaluation and safety: cloned tools can contaminate benchmark splits, bias measurements of generalization across tools, and propagate vulnerable or unsafe implementation patterns across repositories. Our work complements prior studies of tool-using agents by shifting the unit of analysis from the agent to the tool ecosystem, and by measuring code-level reuse among agent-tool repositories at scale.

\subsection{Code Similarity and Cloning}
Code cloning has been extensively studied in software engineering. Early work introduced textual, token-based~\citep{kamiya2002ccfinder}, and syntax-aware methods for detecting duplicated or near-duplicated code fragments, including suffix-based detection and AST-based clone analysis~\citep{baxter1998clone,jiang2007deckard,sajnani2016sourcerercc}. Subsequent surveys and comparative studies categorize clone types and detection techniques across lexical, syntactic, and semantic levels~\citep{roy2007survey,koschke2007survey,bellon2007comparison,roy2009comparison}. This line of work establishes that code similarity can be measured using a range of representations, from raw text and token sets to abstract syntax trees and learned program embeddings.

Prior work also shows that code cloning is not merely redundant implementation. Clones can increase maintenance cost, cause inconsistent bug fixes, and propagate defects across software systems~\citep{juergens2009code,li2006cp}. At the ecosystem level, cloning has been studied in mobile application markets, where repackaged or near-duplicate applications can distort marketplace measurements and introduce security risks~\citep{zhou2012dissecting,zhou2012detecting,crussell2012attack,crussell2013andarwin}. Beyond reliability and security, code cloning also raises provenance and ownership concerns: copied implementations may violate license requirements, obscure attribution, or constitute intellectual-property infringement when redistributed without permission. These concerns are especially relevant in public distribution platforms, where independently listed artifacts may share substantial implementation structure.

Our setting differs from traditional clone-detection studies in both artifact type and motivation. Agent tools combine executable code with model-facing metadata, including tool names, natural-language descriptions, and input schemas. Many tools are also implemented as API wrappers or scaffolded servers, making template reuse especially likely. In this setting, cloning can inflate the apparent diversity of an agent-tool ecosystem, contaminate benchmark splits, propagate vulnerable code patterns, and obscure the provenance of reused implementations. We therefore study cloning at the repository level using complementary lexical and fuzzy-structural similarity metrics. Rather than proposing a new clone detector, our goal is to use interpretable similarity measurements and manual verification to characterize implementation diversity and hidden duplication in agentic-AI tool ecosystems.
\section{Dataset Construction}

We construct a unified dataset of agent tools by aggregating repositories from two major ecosystems: MCP servers and Skills. In this section, we describe the structure of each benchmark, the data collection process, and the preprocessing pipeline used to ensure consistency and quality.

\subsection{Benchmark Structure}
\myparatight{MCP Servers}
An MCP server is a host implementing MCP, which enables LLMs to interact with external systems via executable tools. Each server exposes a collection of tools, where each tool is defined by a function signature, an argument schema, and a natural language description. We collect MCP server from three public marketplaces: MCP.so~\cite{mcpso}, MCP Servers~\cite{mcpservers}, and MCP Market~\cite{mcpmarket}. For each server, we extract both metadata and associated GitHub repositories, including attributes such as server name, developer, descriptions, tool lists, and configuration details. The details for each marketplace are shown in Appendix~\ref{appendix:mcpmarket}.

\myparatight{Skills}
In addition to MCP tools, we collect AI skills from SkillsMP~\cite{skillsmp},
which represent higher-level reusable instruction modules. Unlike MCP tools, which expose atomic functionalities, Skills tools encode structured procedures in natural language, typically implemented as lightweight repositories containing markdown-based instructions. For each Skills tool, we collect metadata including name, developer, description, GitHub URL, and usage-related attributes.

\myparatight{Unified Representation}
To enable comparison across heterogeneous tool ecosystems, we unify MCP servers and Skills at the repository level, using GitHub repositories as the canonical unit of analysis. Specifically, each MCP server (hereafter referred to as an MCP repository) and Skills entry (hereafter referred to as a Skills repository) is mapped to a repository $C_i$. Each repository $C_i$ serves as a container for a distinct set of tools, encompassing both the implementation code and associated metadata (e.g., descriptions, authorship, and usage information). This abstraction allows us to treat tools from different platforms under a common representation. By reducing all artifacts to repository-level objects, we enable direct cross-platform similarity analysis, which forms the basis of our cloning measurement.

\subsection{Basic Dataset Statistics}

Details of our data collection and preprocessing pipeline are provided in Appendix~\ref{appendix:collection_and_preprocess}. We summarize the key properties of the collected dataset, including scale, language distribution, and structural characteristics across MCP repositories and Skills repositories.

\myparatight{Dataset Overview}
After preprocessing, the dataset contains 7,508 MCP repositories and 1,353 Skills repositories. We successfully extracted tools from 4,962 of those MCP repositories. In total, the dataset yields 87,564 MCP tools and 12,447 Skills tools, averaging 17.65 tools per extracted MCP repository and 9.20 tools per Skills repository, as summarized in Table~\ref{tab:stats} in Appendix.

\myparatight{Language and Structural Distribution}
The dataset spans a wide range of programming languages. Python, TypeScript, and JavaScript dominate the ecosystem, reflecting the prevalence of web-based and scripting-oriented tool implementations. Table~\ref{tab:languages} in Appendix shows the distribution of repositories by primary language. The number of tools per repository exhibits a highly skewed distribution, where a small fraction of repositories contain a large number of tools, while the majority implement only a few. 
\section{Tool Metadata Analysis}
\label{sec:metadata-analysis}
We first analyze metadata signals of the collected tool ecosystem before turning to code-level similarity. The goal of this section is not to identify clones from metadata alone, but to characterize whether the apparent scale of the ecosystem is accompanied by diversity in tool interfaces, functionality, and contributors. We analyze three aspects: description length, functionality categories, and developer concentration.

\subsection{Tool Description Length}
\label{sec:description-length}
Tool descriptions are the main natural-language field exposed to LLM agents during tool selection. For MCP servers, we analyze the descriptions of tools collected from the MCP marketplaces. For Skills, we analyze the corresponding tool descriptions collected from the Skills API. We normalize each description by removing formatting artifacts, collapsing repeated whitespace, and tokenizing the resulting text. Empty descriptions are excluded from this analysis.

\begin{figure}[t]
    \centering
    \begin{minipage}[b]{0.4\textwidth}
        \centering
        \includegraphics[width=\linewidth]{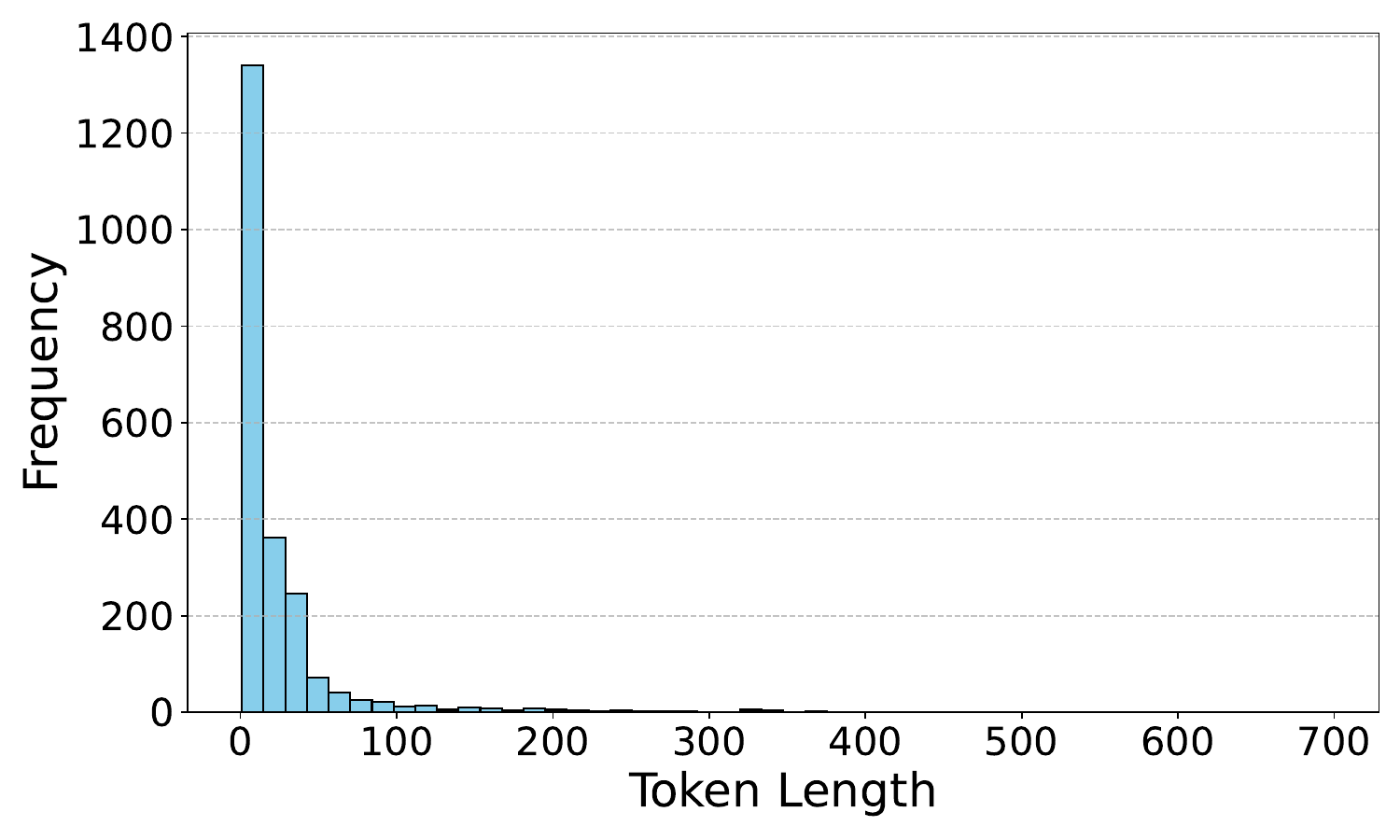}
        \centerline{(a) MCP tools}
    \end{minipage}
    \begin{minipage}[b]{0.4\textwidth}
        \centering
        \includegraphics[width=\linewidth]{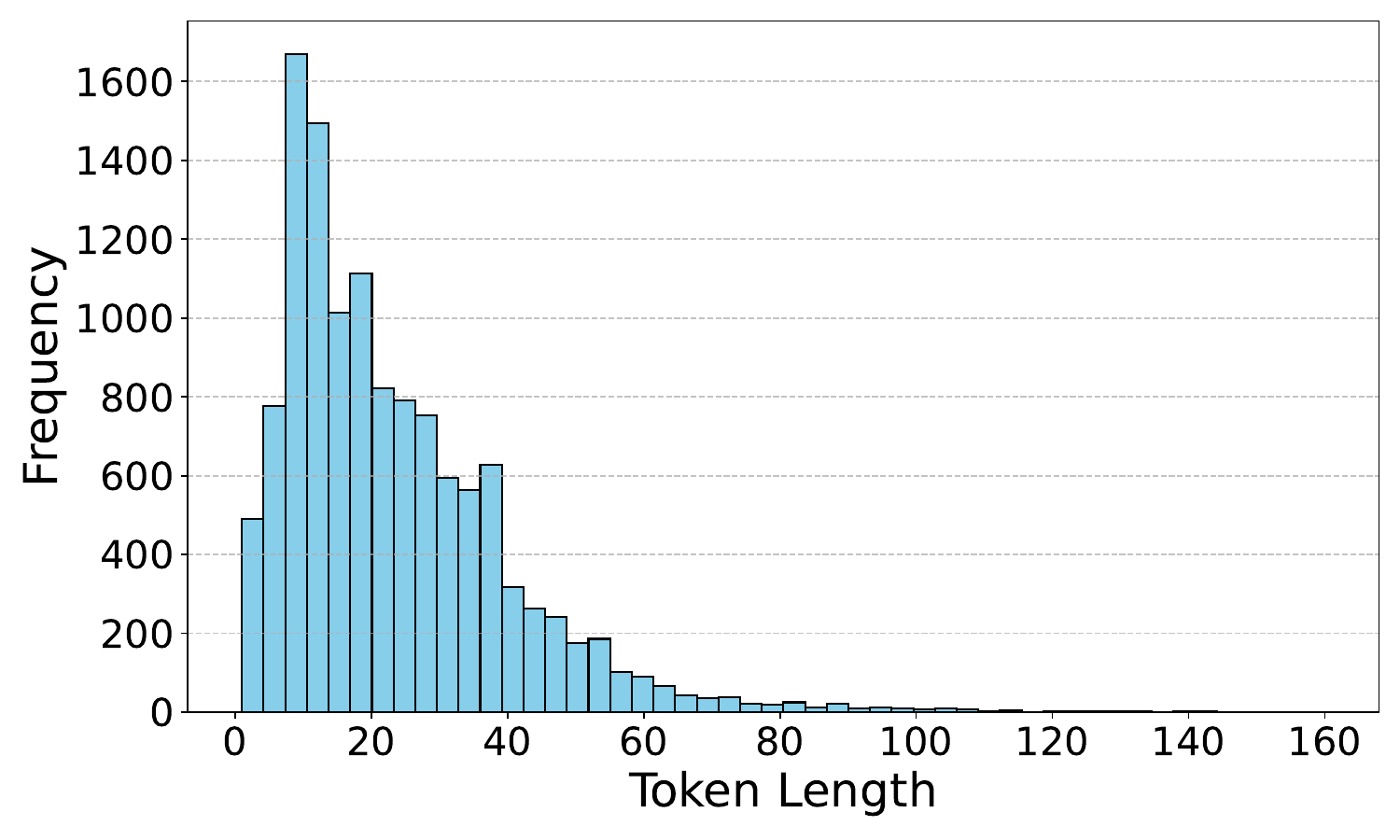}
        \centerline{(b) Skills tools}
    \end{minipage}
    \caption{Description length distributions for MCP and Skills tools.}
    \label{fig:description-length}
\end{figure}

Figure~\ref{fig:description-length} shows the resulting description-length distributions. Both ecosystems exhibit highly skewed distributions. For MCP tools, the median description length is 10 tokens, while the 90th and 95th percentiles are 48 and 96 tokens, respectively. The longest MCP tool description contains 694 tokens. Thus, half of MCP tools are described with no more than a short phrase or sentence, while a small number of tools include much more detailed usage guidance.

Skills exhibit a similar right-skewed pattern but with a narrower range. The median Skills tool description contains 19 tokens, the mean is approximately 23 tokens, and the 90th and 95th percentiles are 44 and 54 tokens. The maximum Skills tool description length is 160 tokens, substantially shorter than the maximum observed for MCP tools.

These results show that model-facing documentation is not standardized across the ecosystem. Some tools expose only minimal descriptions, while others provide detailed instructions. However, description-level variation should not be interpreted as implementation-level diversity. Tools with different descriptions may still share the same scaffold or wrapper code, while tools with similar descriptions may differ in implementation. This motivates our later repository-level analysis, which directly measures code similarity rather than relying on metadata alone.

\subsection{Functionality Categorization}
\label{sec:functionality-categorization}

We next characterize the functional composition of the tool ecosystem collected from the MCP marketplaces or from the Skills API. We define a fixed taxonomy based on recurring tool patterns observed in the dataset: data retrieval, API interaction, file manipulation, database access, code execution, communication, system operations, developer tooling, and other. Each tool is assigned one or more categories based on its name, description, and input schema when available.

To scale the annotation, we use Llama-4-Scout-17B-16E~\citep{meta_llama4_2025} as a classifier with the fixed label set above. The classifier is prompted to select only from the predefined taxonomy and to allow multi-label assignments when a tool spans multiple functions. We manually inspect 50 randomly sampled annotations from MCP tools and 50 from Skills tools, and find that 94\% and 92\%, respectively, are correct under our taxonomy. Most disagreements arise from ambiguous tools that combine API access with data retrieval. The system prompt we use is shown in Appendix~\ref{app:functionality-prompt}.

\begin{figure}[t]
    \centering
    \begin{subfigure}[b]{0.4\textwidth}
        \centering
        \includegraphics[width=\linewidth]{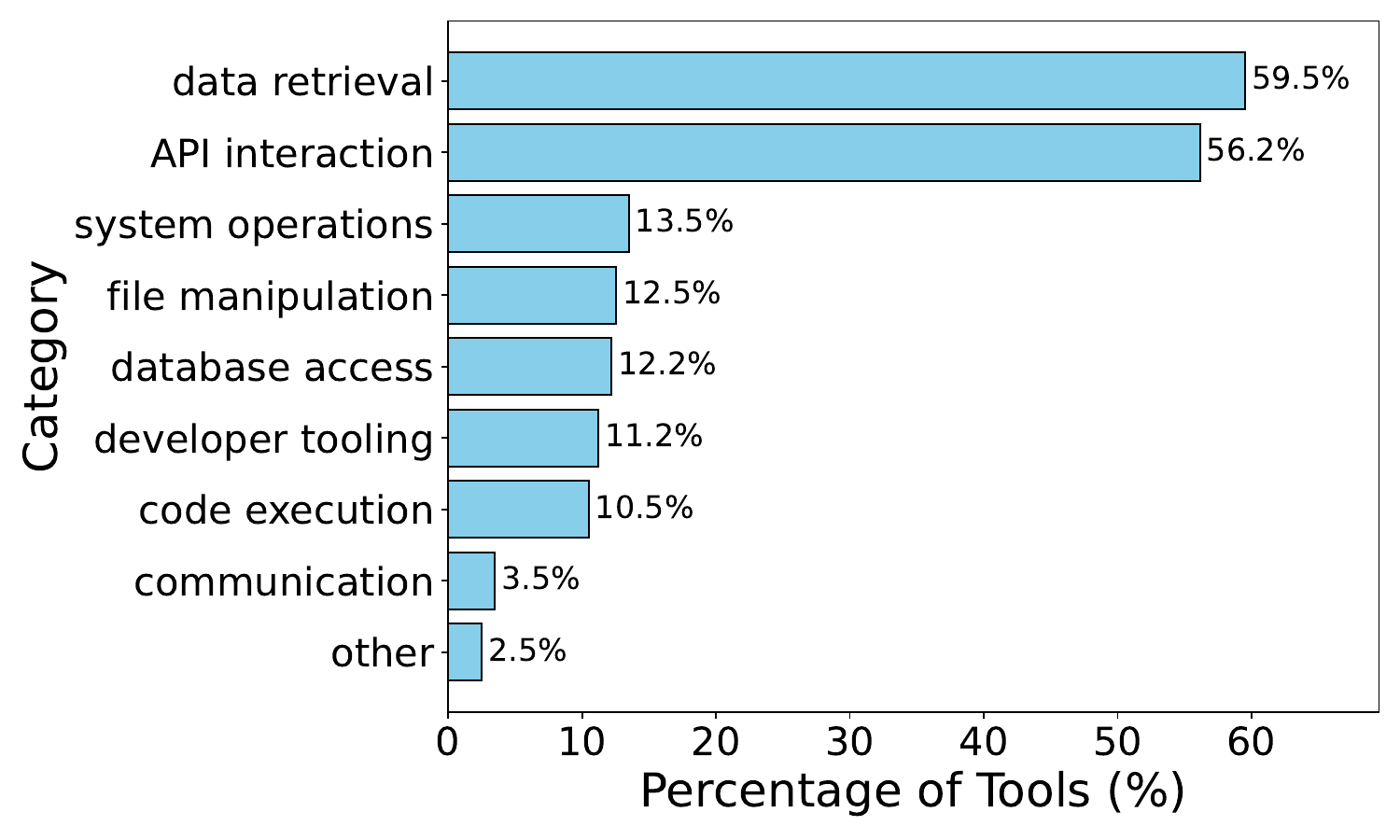}
        \caption{}
        \label{fig:mcp-functionality}
    \end{subfigure}
    \hspace{4mm}
    \begin{subfigure}[b]{0.4\textwidth}
        \centering
        \includegraphics[width=\linewidth]{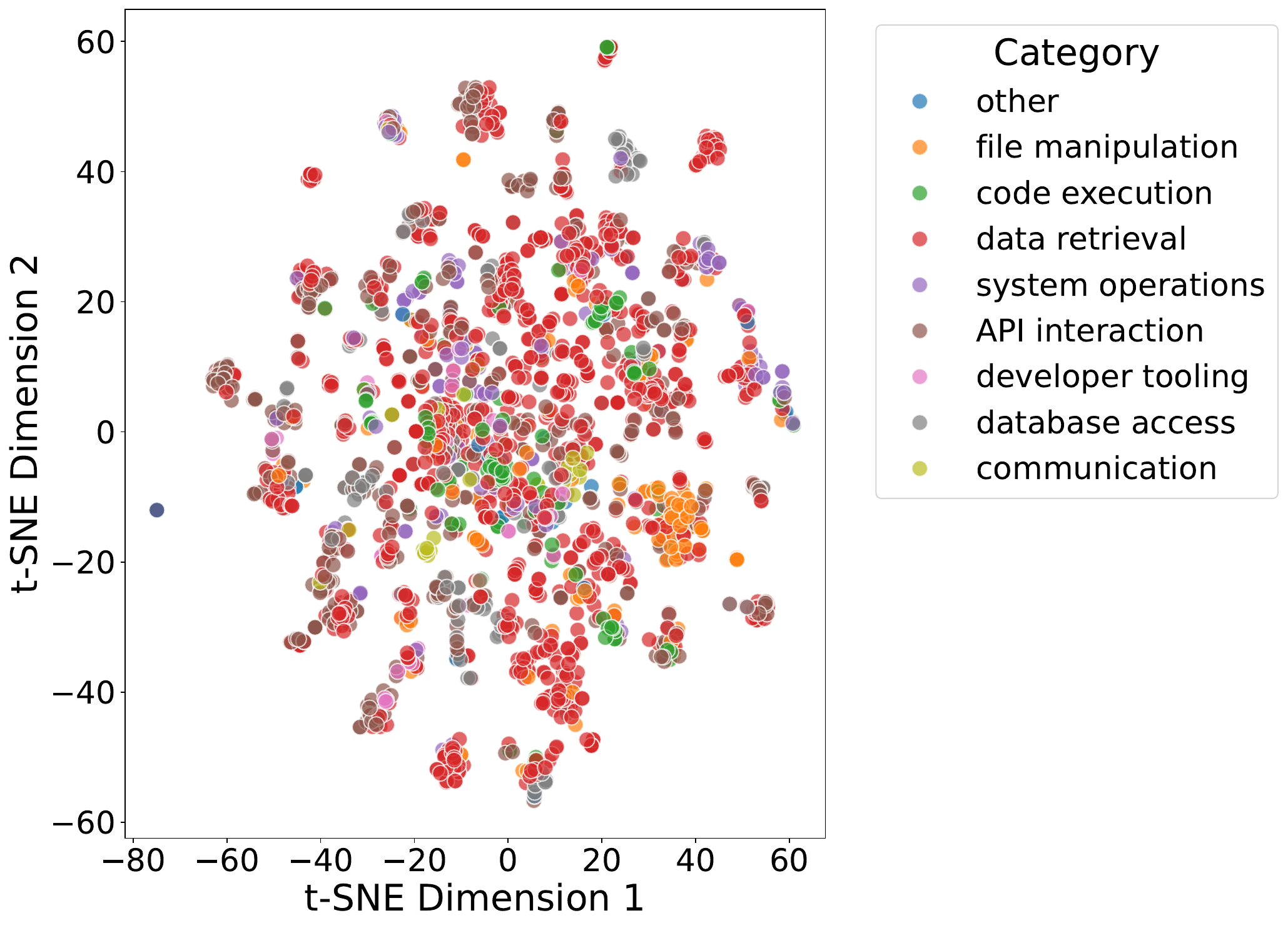}
        \caption{}
        \label{fig:mcp-tsne}
    \end{subfigure}

    \caption{
    Functionality and description-space analysis of MCP tools.
    (a) MCP functionality distribution.
    (b) t-SNE projection of TF-IDF representations for MCP tool descriptions.
    Category percentages may sum to more than 100\% because tools can receive multiple labels.
    The t-SNE plot visualizes metadata-level semantic diversity only; it does not measure implementation diversity.
    }
    \label{fig:mcp-functionality-metadata}
\end{figure}

Figure~\ref{fig:mcp-functionality} and \ref{fig:skills-functionality} in Appendix report the resulting category distributions. MCP tools are concentrated in a small number of categories: data retrieval and API interaction together account for 76.6\% of tools, followed by system operations at 13.5\% and file manipulation at 12.5\%. Developer tooling and code execution account for 11.2\% and 10.5\%, respectively. Skills tools exhibit a different but similarly concentrated distribution: developer tooling accounts for 59.1\% of skills, followed by code execution at 24.1\%.

\myparatight{Description-space visualization}
We further visualize the semantic structure induced by model-facing descriptions. For each MCP or Skills tool, we construct a TF-IDF representation from its name and natural-language description, and project the representations into two dimensions using t-SNE. 

Figure~\ref{fig:mcp-tsne} and Figure~\ref{fig:skills-tsne} in Appendix show that both MCP and Skills tools occupy diverse regions in description space, suggesting substantial apparent diversity in model-facing metadata. However, categories are not cleanly separated, and this visualization only reflects metadata-level semantic variation. It does not determine whether repositories correspond to independent implementations. We therefore use this analysis only to characterize the model-facing surface of the ecosystem, and rely on repository-level code similarity and manual verification in Section~\ref{sec:code-cloning} to measure implementation-level duplication.

Overall, the functionality and description-space analyses show that the ecosystem contains a broad range of model-facing intents, but that these intents are concentrated around recurring task patterns. This concentration is important for interpreting code cloning. Many tools implement recurring integration tasks, especially data access, API interaction, developer tooling, and code execution. Such tools often share similar implementation components, including authentication handling, request construction, response parsing, schema registration, and error handling. Functionality concentration therefore does not by itself prove cloning, but it explains why template-derived or wrapper-based reuse is likely in this ecosystem and motivates the repository-level clone analysis in Section~\ref{sec:code-cloning}.

\subsection{Developer Analysis}
\label{sec:developer-analysis}
Building upon our analysis of model-facing marketplace metadata, we transition to the extracted tools to analyze the distribution of tool development across contributors. Specifically, we quantify the number of repositories and individual tools, derived via our preprocessing pipeline, associated with that developer. This allows us to measure whether the ecosystem is broadly distributed across contributors or dominated by a small number of highly active developers.

\begin{figure}[htp]
    \centering
    \begin{minipage}[b]{0.4\textwidth}
        \centering
        \includegraphics[width=\linewidth]{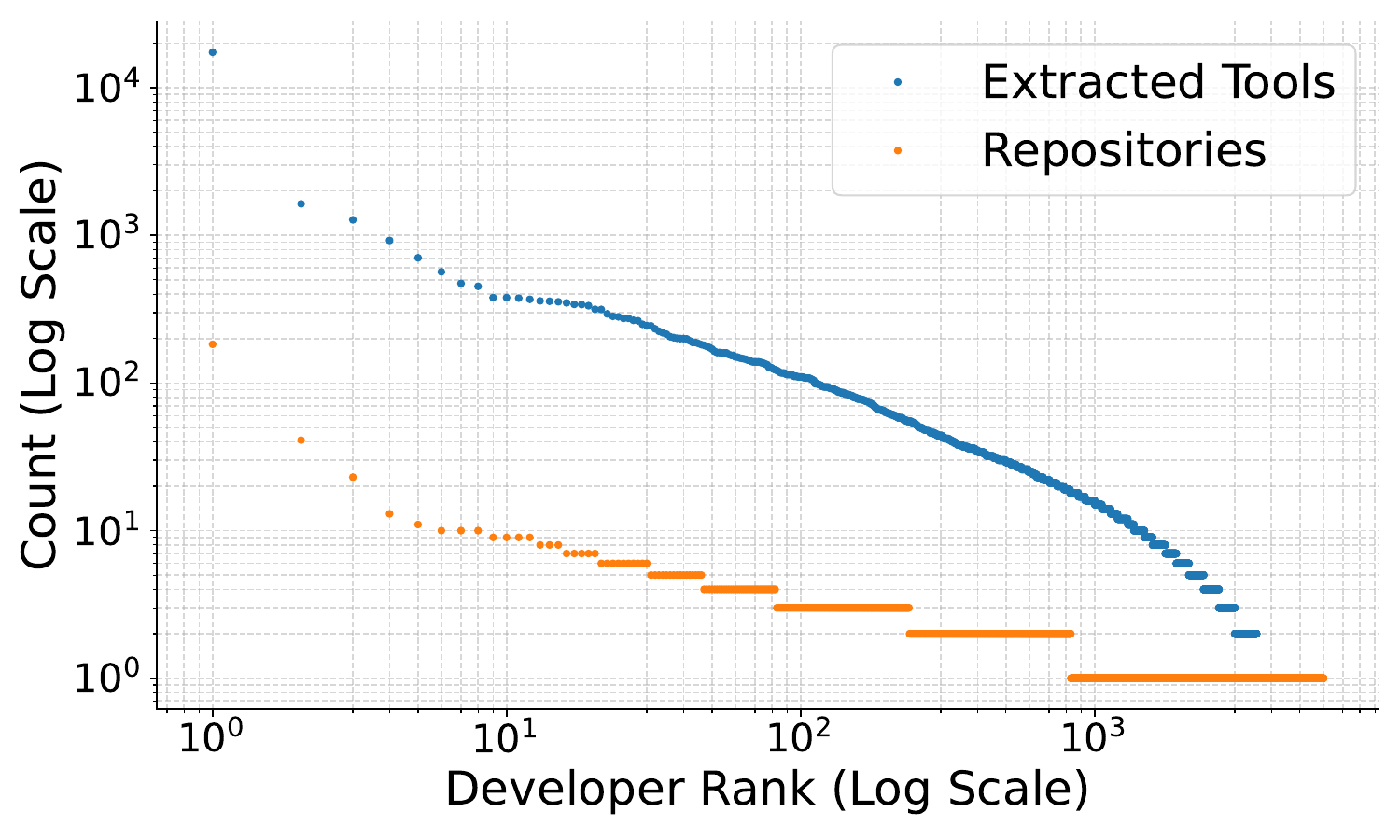}
        \centerline{(a) MCP developers}
    \end{minipage}
    \begin{minipage}[b]{0.4\textwidth}
        \centering
        \includegraphics[width=\linewidth]{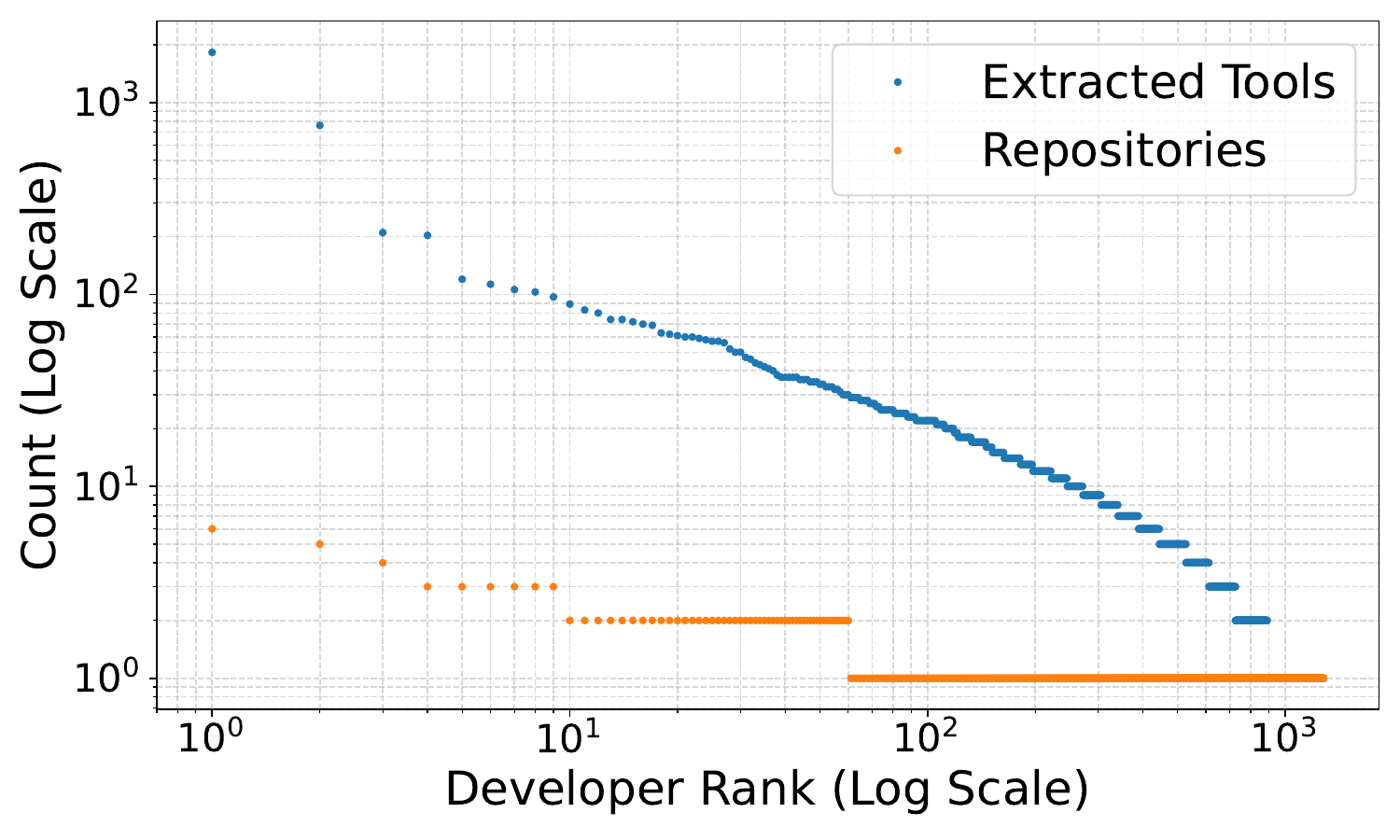}
        \centerline{(b) Skills developers}
    \end{minipage}
    \caption{Developer contribution distributions in the MCP and Skills ecosystems.}
    \label{fig:developer-distribution}
\end{figure}

Figure~\ref{fig:developer-distribution} and Table~\ref{tab:developer-concentration} in Appendix show that tool contributions are highly concentrated. In the MCP ecosystem, the top 10 developers account for 27.6\% of extracted tools while owning only 2.6\% of repositories; the top 50 developers account for 39.3\% of tools and 4.1\% of repositories. This indicates that a small number of developers publish repositories that expose many tools. The Skills ecosystem shows a similar pattern: the top 10 developers account for 29.1\% of tools, and the top 50 account for 45.5\%.

Developer concentration matters for clone analysis because similarity has different interpretations depending on authorship. Similar repositories from the same developer may reflect intentional self-reuse, shared project templates, or maintained variants within a tool family. In contrast, high similarity across different developers provides stronger evidence of ecosystem-level propagation, such as copied implementations or shared templates spreading across independently listed repositories. For this reason, our main code-cloning analysis excludes same-developer repository pairs when estimating cross-developer clone prevalence.

Overall, the metadata analysis reveals three properties of the ecosystem: tool descriptions are highly variable, functionality is concentrated around recurring integration tasks, and contributions are dominated by a small number of active developers. These findings suggest that raw tool counts alone are insufficient for measuring ecosystem diversity. They motivate the repository-level similarity analysis in Section~\ref{sec:code-cloning}, where we directly test whether apparent tool diversity corresponds to independent implementations.

\section{Code Cloning Analysis}
\label{sec:code-cloning}
We next analyze whether repositories in the collected agent-tool ecosystem share substantial implementation structure. Our goal is not only to identify individual near-duplicate pairs, but also to measure whether cloning is an isolated phenomenon or a pervasive source of hidden duplication. We proceed in four steps. We first normalize repositories and characterize their code size, then compute repository-level similarity using two complementary metrics, quantify the prevalence of high-similarity candidate pairs, and finally manually verify sampled pairs to calibrate how often high similarity corresponds to true cloning.

\subsection{Code Size Distribution}
\label{sec:code-size}
\myparatight{Repository normalization}
For each repository, we recursively extract source files and construct a normalized repository-level representation. We exclude dependency directories, generated artifacts, and common build outputs, including \texttt{.git}, \texttt{node\_modules}, \texttt{dist}, \texttt{build}, and \texttt{\_\_pycache\_\_}. We also remove binary and archive formats such as \texttt{.png}, \texttt{.jpg}, \texttt{.gif}, \texttt{.zip}, \texttt{.tar}, and \texttt{.gz}. The remaining source files are concatenated after comment removal, whitespace normalization, and lowercasing.

\myparatight{Size distribution}
Figure~\ref{fig:code-size} in Appendix shows the distribution of repository sizes measured by normalized source tokens. The distribution for MCP repositories is heavy-tailed: many repositories contain lightweight tool wrappers, while a smaller number contain substantially larger server implementations or multi-tool packages. The distribution for Skills repositories follows a more normal curve, indicating standard complexity across most skill packages rather than large outliers. To avoid unstable similarity estimates from trivial repositories, we exclude repositories with fewer than 50 normalized tokens. This removes incomplete or non-informative repositories while preserving the large majority of tool implementations.

\subsection{Repository Similarity Metrics}
\label{sec:similarity-metrics}

\myparatight{Comparison setting}
We compute pairwise similarity over three groups: MCP-to-MCP repository pairs, Skills-to-Skills repository pairs, and cross-domain MCP-to-Skills pairs. Unless otherwise stated, we exclude repository pairs authored by the same developer when estimating ecosystem-level cloning. Same-developer similarity is analyzed separately because it may reflect intentional self-reuse, shared project templates, or maintained tool families rather than cross-developer reuse.

\subsection{Similarity Distributions}
\label{sec:similarity-results}

\begin{figure}[t]
    \centering

    \begin{minipage}[t]{0.32\textwidth}
        \centering
        \includegraphics[width=\linewidth]{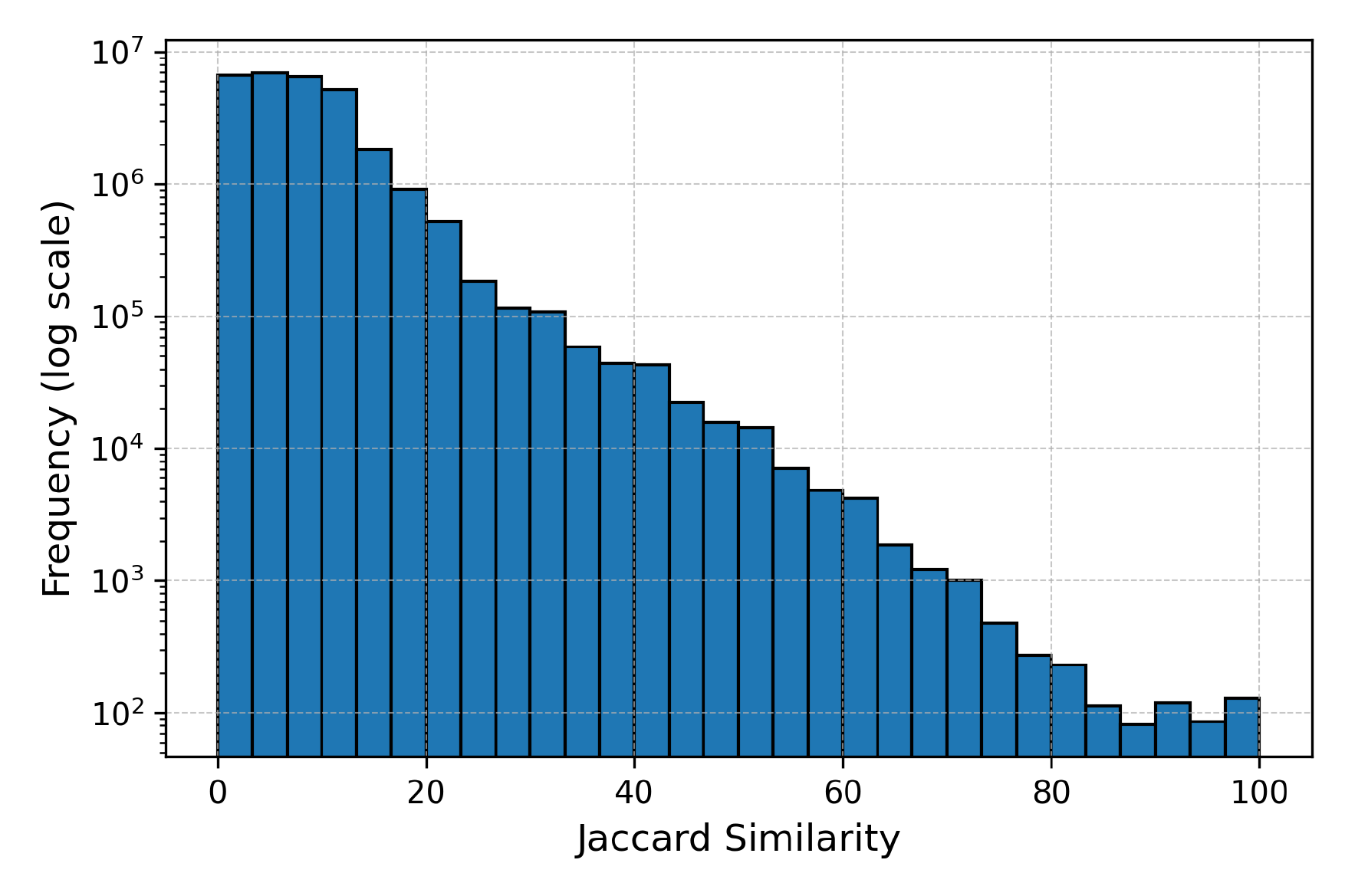}
        \centerline{(a) MCP--MCP Jaccard}
    \end{minipage}
    \hfill
    \begin{minipage}[t]{0.32\textwidth}
        \centering
        \includegraphics[width=\linewidth]{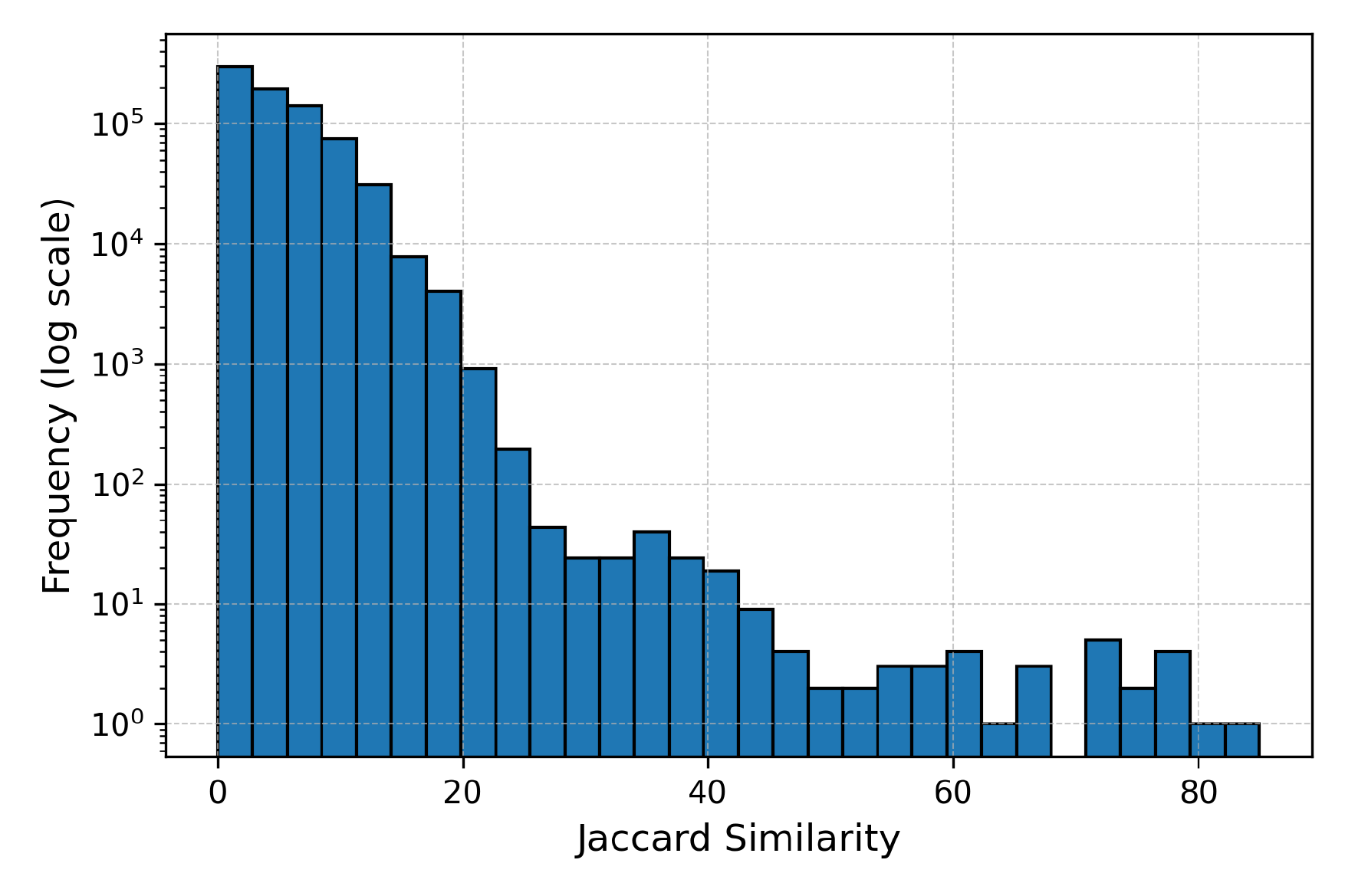}
        \centerline{(b) Skills--Skills Jaccard}
    \end{minipage}
    \hfill
    \begin{minipage}[t]{0.32\textwidth}
        \centering
        \includegraphics[width=\linewidth]{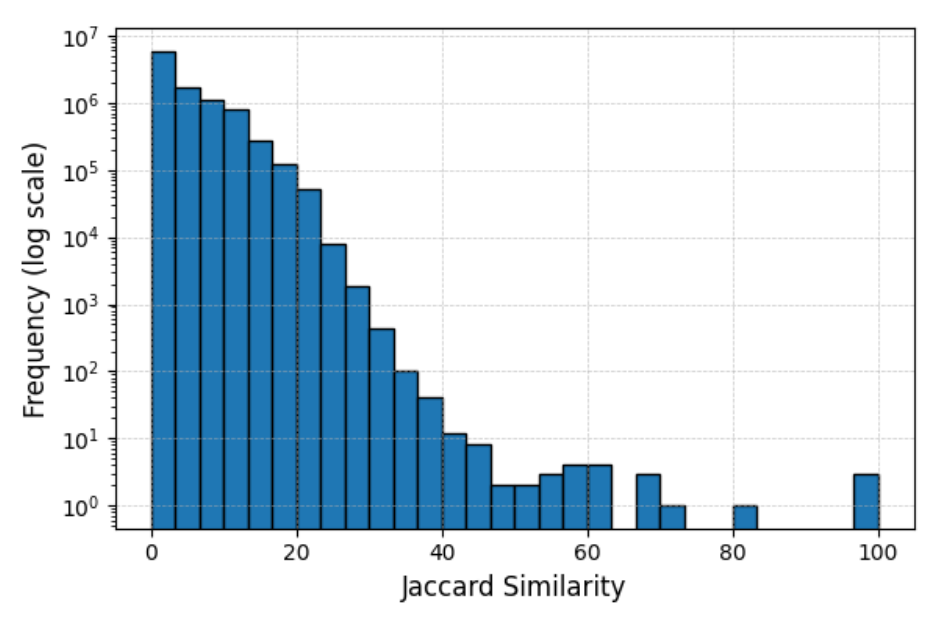}
        \centerline{(c) MCP--Skills Jaccard}
    \end{minipage}

    \vspace{0.8em}

    \begin{minipage}[t]{0.32\textwidth}
        \centering
        \includegraphics[width=\linewidth]{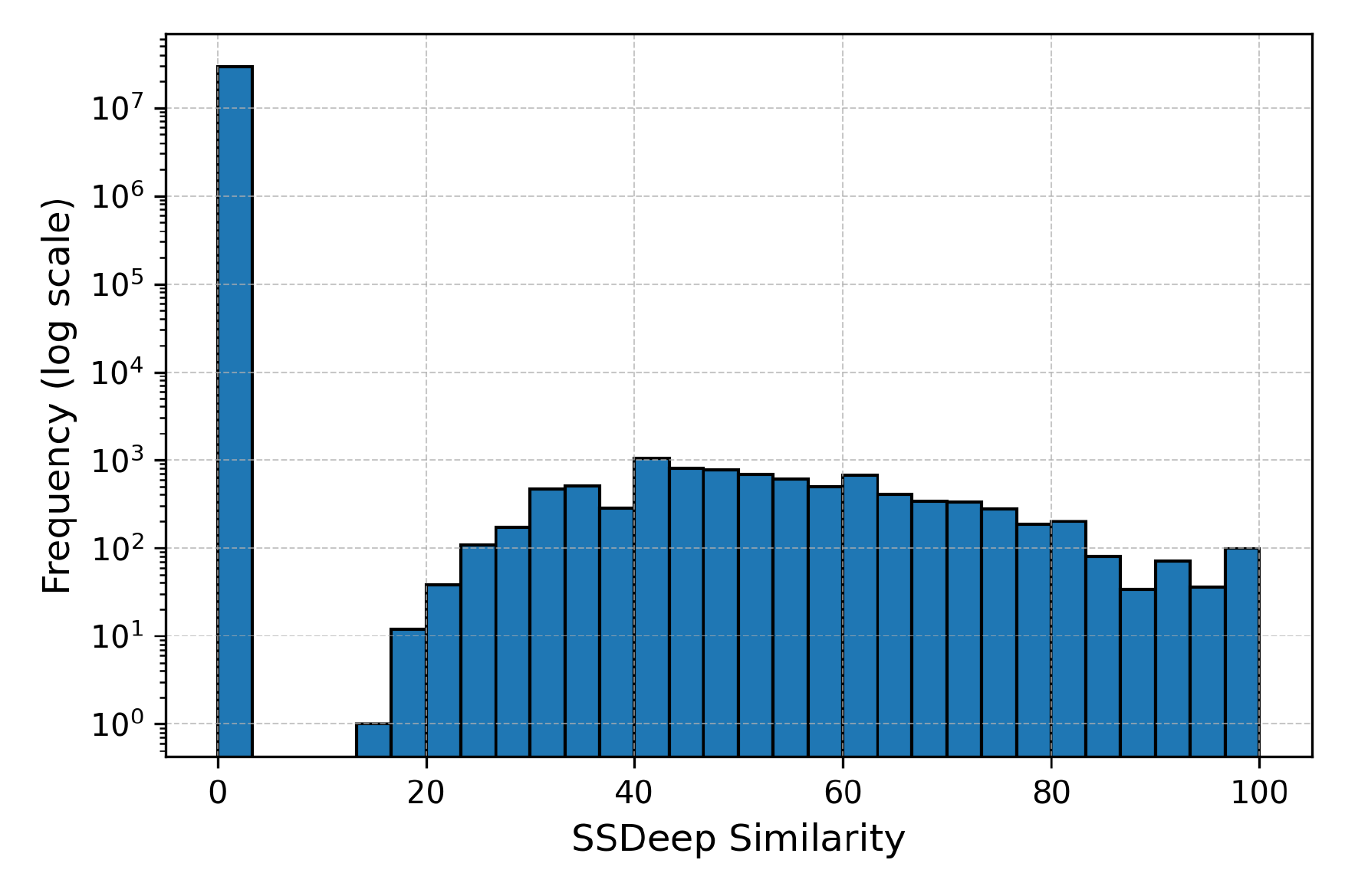}
        \centerline{(d) MCP--MCP \texttt{ssdeep}}
    \end{minipage}
    \hfill
    \begin{minipage}[t]{0.32\textwidth}
        \centering
        \includegraphics[width=\linewidth]{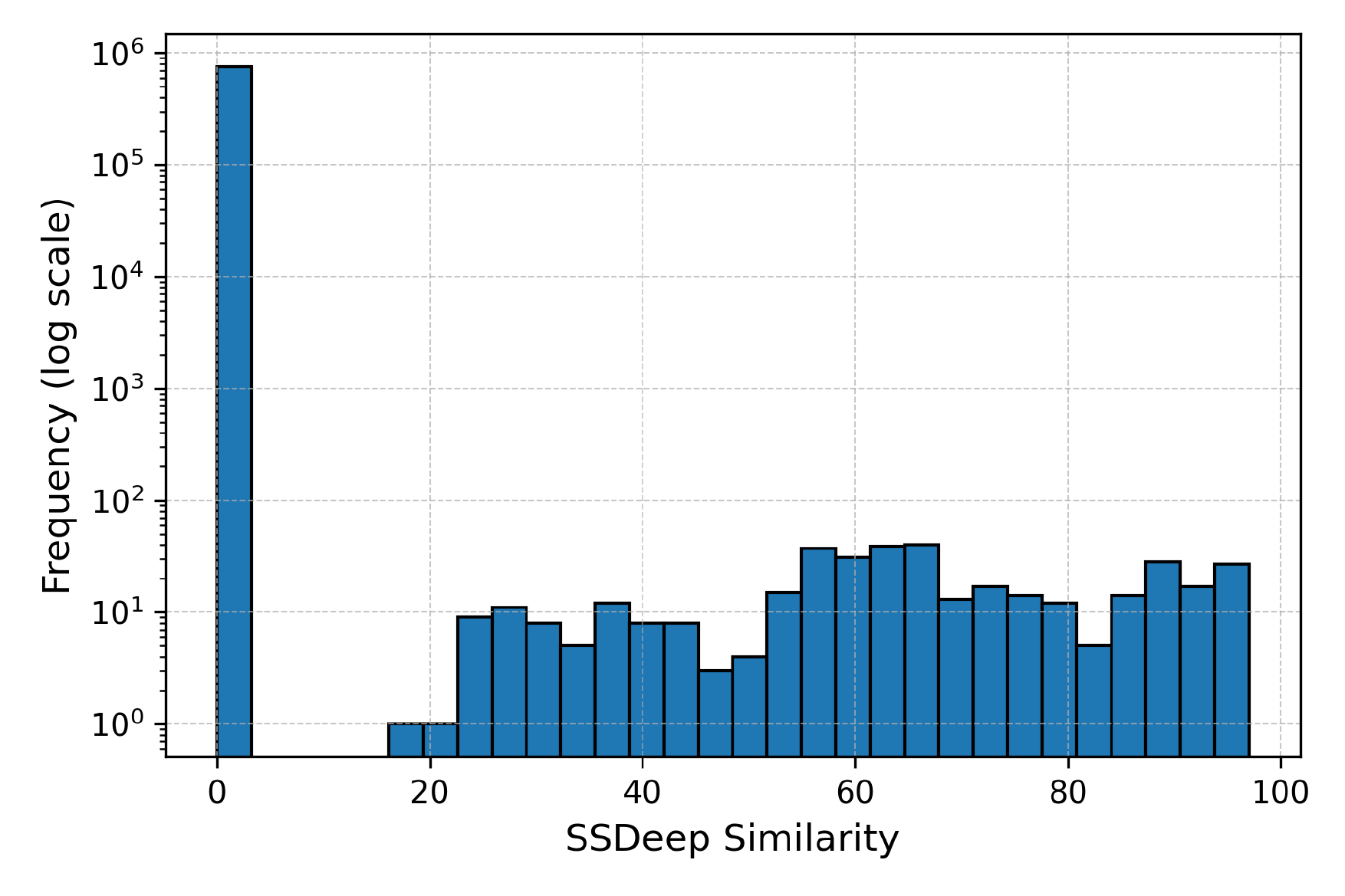}
        \centerline{(e) Skills--Skills \texttt{ssdeep}}
    \end{minipage}
    \hfill
    \begin{minipage}[t]{0.32\textwidth}
        \centering
        \includegraphics[width=\linewidth]{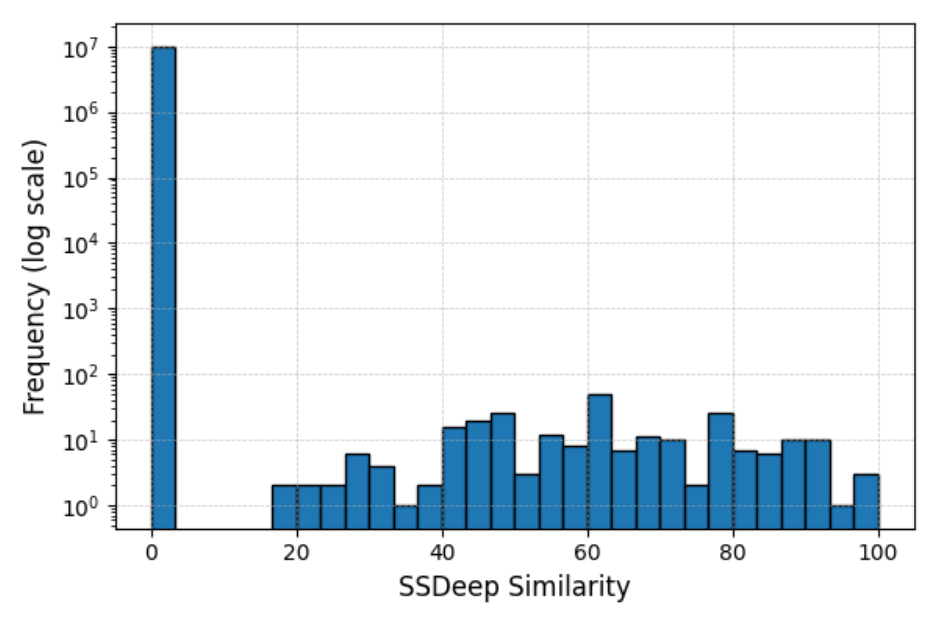}
        \centerline{(f) MCP--Skills \texttt{ssdeep}}
    \end{minipage}

    \caption{
    Pairwise repository similarity distributions across three comparison groups.
    The top row shows Jaccard similarity distributions, and the bottom row shows \texttt{ssdeep} similarity distributions.
    }
    \label{fig:similarity-distributions}
\end{figure}

Using the similarity metrics outlined in Appendix~\ref{appendix:similarity_metrics}, we compute pairwise Jaccard and \texttt{ssdeep} similarity scores for the three comparison groups described above. Figure~\ref{fig:similarity-distributions} shows the resulting score distributions. We plot the two metrics separately because they capture different forms of reuse: Jaccard measures token-level overlap, whereas \texttt{ssdeep} measures preservation of contiguous code regions.

For Jaccard similarity, most repository pairs fall into low-score regions, indicating limited token-level overlap across the majority of the ecosystem. However, each comparison group also contains a high-score tail of candidate pairs with substantial lexical reuse. These pairs may correspond to near-duplicates, lightly modified forks, or repositories instantiated from common templates, although the metric alone cannot distinguish true clones from false positives such as shared library names, common API vocabulary, or repeated framework code.

The \texttt{ssdeep} distributions provide a complementary view. Most pairs again receive scores near zero, while non-zero high-score regions identify candidate pairs that preserve longer contiguous code regions. Such pairs may share implementation scaffolds or copied code blocks even when token overlap is reduced by renaming or localized edits. Since high \texttt{ssdeep} similarity can also arise from shared boilerplate or framework conventions, we use these scores as candidate signals rather than clone labels.

\subsection{Prevalence of High-Similarity Candidates}
\label{sec:candidate-prevalence}

To assess whether cloning is isolated or widespread, we count high-similarity candidate pairs for each metric and comparison group. Beyond pair counts, we report the number of unique repositories involved and the size of the largest candidate clone cluster, where a cluster is defined as repositories connected through one or more high-similarity pairwise relationships. These statistics indicate whether high-similarity reuse is limited to isolated pairs or spans larger groups of repositories.

Table~\ref{tab:clone-candidate-prevalence} in Appendix shows that high-similarity candidates are not confined to a few isolated repositories. They appear across MCP-to-MCP, Skills-to-Skills, and cross-domain MCP-to-Skills comparisons, involving 764 and 648 repositories respectively for jaccard and \texttt{ssdeep} in total and forming candidate clusters of up to 38 repositories. This indicates that implementation reuse is a broad ecosystem-level phenomenon rather than a small number of accidental duplicate pairs. The next subsection manually verifies sampled candidates to determine how often these high-similarity regions correspond to true clones.

\subsection{Manual Verification and Score Calibration}
\label{sec:manual-verification}

Similarity scores provide candidate clone pairs, but do not by themselves establish cloning. High similarity may be caused by shared dependencies, common framework scaffolds, package manifests, or domain-specific API vocabulary. We therefore manually verify sampled repository pairs from different similarity ranges.

For each metric and comparison group, we divide repository pairs into score buckets and randomly sample pairs from each bucket for inspection. Annotators follow a fixed verification rubric, with the full rubric provided in Appendix~\ref{app:verification-rubric}. They first compare repository layout and file organization, then inspect core source files to determine whether the repositories share non-trivial implementation logic. Similarity caused only by dependency manifests, generated files, framework initialization, or generic boilerplate is discounted. A pair is labeled as \texttt{clone} only if the overlap extends to substantive implementation logic, such as tool registration, API handling, request construction, response parsing, authentication flow, file-system operations, command execution, or equivalent tool behavior. Minor renaming, configuration edits, or localized refactoring do not disqualify a clone label; substantial architectural divergence results in a \texttt{non-clone} label.

\begin{table}[t]
\centering
\caption{Manual verification results by comparison group and similarity bucket. Clone rate is computed as verified clones divided by sampled pairs.}
\label{tab:manual-verification}
\resizebox{\textwidth}{!}{%
\begin{tabular}{ccc}
\toprule
\textbf{MCP--MCP} & \textbf{Skills--Skills} & \textbf{MCP--Skills} \\
\midrule

\begin{tabular}{lllrr}
\toprule
Metric & Bucket & Total Pairs & Clone/Samp. & Proportion (95\% CI) \\
\midrule
Jaccard & 0--20   & 28034862 & 0/20 & 0.00 (0.00--0.16) \\
Jaccard & 20--40  & 1033069   & 0/20 & 0.00 (0.00--0.16) \\
Jaccard & 40--60  & 107481    & 5/20 & 0.25 (0.11--0.47) \\
Jaccard & 60--80  & 9054      & 6/20 & 0.30 (0.15--0.52) \\
Jaccard & 80--100 & 758       & 12/20 & 0.60 (0.39--0.78) \\
\midrule
\texttt{ssdeep} & 0--20   & 29176526 & 0/20 & 0.00 (0.00--0.16) \\
\texttt{ssdeep} & 20--40  & 1584     & 1/20 & 0.05 (0.01--0.24) \\
\texttt{ssdeep} & 40--60  & 4403     & 3/20 & 0.15 (0.05--0.36) \\
\texttt{ssdeep} & 60--80  & 2194     & 9/20 & 0.45 (0.26--0.66) \\
\texttt{ssdeep} & 80--100 & 517      & 17/20 & 0.85 (0.64--0.95) \\
\bottomrule
\end{tabular}
&
\begin{tabular}{lllrr}
\toprule
Metric & Bucket & Total Pairs & Clone/Samp. & Proportion (95\% CI) \\
\midrule
Jaccard & 0--20   & 749504  & 0/20 & 0.00 (0.00--0.16) \\
Jaccard & 20--40  & 1263    & 0/20 & 0.00 (0.00--0.16) \\
Jaccard & 40--60  & 42      & 2/20 & 0.10 (0.03--0.30) \\
Jaccard & 60--80  & 19      & 8/19 & 0.42 (0.23--0.64) \\
Jaccard & 80--100 & 2       & 2/2  & 1.00 (0.34--1.00) \\
\midrule
\texttt{ssdeep} & 0--20   & 750452 & 0/20  & 0.00 (0.00--0.16) \\
\texttt{ssdeep} & 20--40  & 46     & 0/20  & 0.00 (0.00--0.16) \\
\texttt{ssdeep} & 40--60  & 75     & 3/20  & 0.15 (0.05--0.36) \\
\texttt{ssdeep} & 60--80  & 163    & 8/20  & 0.40 (0.22--0.61) \\
\texttt{ssdeep} & 80--100 & 94     & 15/20 & 0.75 (0.53--0.89) \\
\bottomrule
\end{tabular}
&
\begin{tabular}{lllrr}
\toprule
Metric & Bucket & Total Pairs & Clone/Samp. & Proportion (95\% CI) \\
\midrule
Jaccard & 0--20   & 9964270  & 0/20 & 0.00 (0.00--0.16) \\
Jaccard & 20--40  & 62827    & 0/20 & 0.00 (0.00--0.16) \\
Jaccard & 40--60  & 31       & 0/20 & 0.00 (0.00--0.16) \\
Jaccard & 60--80  & 8        & 4/8  & 0.50 (0.22--0.78) \\
Jaccard & 80--100 & 4        & 1/4  & 0.25 (0.05--0.70) \\
\midrule
\texttt{ssdeep} & 0--20   & 10026897 & 0/20 & 0.00 (0.00--0.16) \\
\texttt{ssdeep} & 20--40  & 17       & 0/17 & 0.00 (0.00--0.18) \\
\texttt{ssdeep} & 40--60  & 84       & 1/20 & 0.05 (0.01--0.24) \\
\texttt{ssdeep} & 60--80  & 105      & 3/20 & 0.15 (0.05--0.36) \\
\texttt{ssdeep} & 80--100 & 37       & 5/20 & 0.25 (0.11--0.47) \\
\bottomrule
\end{tabular}
\\
\bottomrule
\end{tabular}%
}
\end{table}

Table~\ref{tab:manual-verification} reports manual verification results across similarity buckets and comparison groups. The clone rate increases with similarity in all three settings, showing that high-score regions are consistently enriched for true implementation reuse. The trend is strongest for MCP--MCP pairs, followed by Skills--Skills pairs, while MCP--Skills pairs exhibit lower but nonzero clone rates. This pattern suggests that reuse is most common within each artifact type, but also appears across ecosystem boundaries.

We report 95\% Wilson confidence intervals to account for uncertainty from finite manual samples; details are provided in Appendix~\ref{app:wilson-ci}. Buckets with fewer than 20 available candidate pairs are fully inspected and reported with their actual sample sizes. Together with the prevalence statistics in Table~\ref{tab:clone-candidate-prevalence}, these results show that tool cloning is both widespread and practically significant: high-similarity regions involve many repositories, and the highest-score buckets contain a large fraction of verified clones.

\subsection{Implications for Tool Diversity}
\label{sec:verified-clone-implications}
These results have direct implications for measuring tool-ecosystem diversity. Repository and tool counts can substantially overstate effective diversity when high-similarity regions contain many verified clones. A marketplace may list many tools, yet a meaningful fraction of them may be near-duplicates, lightly modified variants, or implementations derived from shared scaffolds.

The results also show why raw similarity scores should not be treated as definitive clone labels. Some high-scoring pairs are false positives caused by shared boilerplate or framework conventions, while some lower-scoring pairs may still share implementation structure that is partially obscured by edits. Manual verification therefore serves as a calibration step that estimates the precision of each score range and prevents overclaiming from metric values alone.

For benchmark construction, random repository-level splits may place similar implementations across train and test sets, causing models to appear to generalize across tools while encountering familiar code patterns. For security and provenance auditing, clone-aware grouping can help identify sets of repositories that should be analyzed together rather than treated as independent tools. This is especially important when copied implementations carry vulnerabilities, unsafe patterns, or unclear attribution and license provenance.
\section{Conclusion and Future Work}
We present a large-scale measurement study of tool cloning in agentic-AI ecosystems. We curate a unified dataset of MCP and Skills repositories, measure repository-level lexical and fuzzy-structural similarity, and manually verify sampled pairs across similarity-score buckets. Our results show that high-similarity regions are consistently enriched for true clones across MCP--MCP, Skills--Skills, and MCP--Skills comparisons, indicating that cloning is a pervasive source of hidden duplication in the ecosystem. These findings suggest that raw repository and tool counts are insufficient for assessing ecosystem diversity, and that implementation provenance should be considered when constructing agent-tool datasets and benchmarks. Future work includes similarity-aware de-duplication for benchmark construction, semantic clone detection for agent tools, and measuring the impact of cloning on downstream agent evaluation.

\bibliography{refs}
\bibliographystyle{plainnat}

\clearpage
\appendix
\section{Appendix}

\subsection{Details of MCP Marketplaces}\label{appendix:mcpmarket}
\myparatight{MCP.So}
The MCP.So platform organizes servers across paginated listings. We iteratively traversed these pages and extracted the following attributes for each server: Server Name, Server URL, Developer Name, Short Description, Overview, GitHub URL, Server Configuration, Tools, and Comments.

\myparatight{MCP Servers}
The MCP Servers platform presents servers through a paginated interface navigated via a next-page mechanism. We automated traversal across all pages and collected the following attributes per server: Server Name, Server URL, Developer Name, Description, Overview, Related Servers, GitHub URL, and Tools.

\myparatight{MCP Market}
MCP Market uses an infinite scrolling interface. We simulated scrolling to enumerate all visible servers and subsequently scraped individual server pages. For each server, we collected: Server Name, Server URL, Developer Name, Short Description, About, Features, Use Cases, FAQs, GitHub URL, Tools, and Related Servers. We note that MCP Market enforces rate limits; therefore, our dataset includes a partial but representative subset of servers due to time and access constraints.

\subsection{Data Collection and Preprocessing}\label{appendix:collection_and_preprocess}
\myparatight{Data Collection}
We implement automated scraping pipelines (Python + Selenium) to traverse marketplace listings and extract structured metadata. For MCP marketplaces, we handle paginated and infinite-scroll interfaces to ensure broad coverage. SkillsMP data is collected via its public API.
Entries from multiple sources are merged based on GitHub repository URLs. Duplicate entries across platforms are consolidated, and metadata fields are aggregated to form a unified record for each repository.

\myparatight{Filtering and Cleaning}
Due to heterogeneity in repository structures, not all MCP repositories yield extractable tool definitions. We apply filtering and normalization steps to ensure consistency. Specifically, we apply the following cleaning processes to ensure data quality:
(i) removing repositories with invalid or inaccessible URLs,  
(ii) excluding non-executable or auxiliary files (e.g., images, archives), and  
(iii) filtering repositories with insufficient content (less than 50 tokens after preprocessing).

\myparatight{Language Identification and Tool Extraction}
We query the GitHub API to obtain language statistics and define the primary programming language as the one with the highest byte count (excluding markup languages such as HTML, CSS, JSON, and YAML). Each repository is cloned locally, and we apply language-specific parsing (AST-based when possible, regex as fallback) to extract tool definitions and estimate the number of tools implemented per repository. Furthermore, each repository is augmented with additional attributes including:
\texttt{languages}, \texttt{primary\_language}, \texttt{local\_repo\_path}, and \texttt{tool\_count}.

\subsection{System Prompt for Functionality Categorization}
\label{app:functionality-prompt}

For functionality categorization, we use Llama-4-Scout-17B-16E as a multi-label classifier. Each input consists of a tool or skill name, its natural-language description, and its input schema when available. The model is instructed to select only from our predefined taxonomy and to return a strict JSON object containing the selected categories and a short explanation. We use deterministic decoding and discard or re-run outputs that cannot be parsed as valid JSON. The full system prompt is shown below.

\begin{Verbatim}[breaklines=true, breakanywhere=true, fontsize=\small]
"You are a tool classification engine. Your task is to categorize a software tool based on its name, description, and input schema (args).

Assign one or more categories from the provided taxonomy. 
- "data retrieval": Fetching data from the web, searching, or reading non-local APIs solely to get information.
- "API interaction": Interacting with external web services (GET/POST) where the primary action is transacting or mutating state.
- "file manipulation": Reading, writing, or editing local files.
- "database access": Querying or modifying structured databases (SQL, NoSQL).
- "code execution": Running, evaluating, or compiling code (Python, bash, etc.).
- "communication": Sending emails, Slack messages, or other human-to-human communications.
- "system operations": OS-level tasks, terminal commands, managing processes or hardware.
- "developer tooling": Git operations, linting, debugging, or IDE integrations.
- "other": Does not fit any of the above.

CRITICAL: You may ONLY use the following categories. DO NOT invent new categories under any circumstances.

If a tool spans multiple functions (e.g., an API that fetches a file), assign all relevant categories.

You MUST return your output in strict JSON format with exactly two keys:
1. "categories": A list containing one or more of the valid categories.
2. "reasoning": A string explaining your reasoning.

Example response:
{
  "categories": ["data retrieval", "API interaction"],
  "reasoning": "The tool retrieves data from an external API endpoint."
}"
\end{Verbatim}

\subsection{Similarity Metrics}\label{appendix:similarity_metrics}
We compute two repository-level similarity scores. First, we measure lexical overlap using Jaccard similarity over normalized token sets. Let \(T_i\) be the set of tokens extracted from repository \(C_i\). For two repositories \(C_i\) and \(C_j\), we define
\[
J(C_i, C_j) = \frac{|T_i \cap T_j|}{|T_i \cup T_j|}.
\]
We report this score as a percentage in \([0,100]\). Jaccard similarity captures exact token reuse and is most useful for identifying near-identical or lightly modified implementations.

Second, we compute fuzzy structural similarity using \texttt{ssdeep}, a context-triggered piecewise hashing method for detecting near-duplicate byte sequences under small edits~\citep{kornblum2006identifying}. Given a repository-level byte sequence \(B_i\), \texttt{ssdeep} produces a fuzzy hash signature \(H(C_i)\); comparing two signatures yields
\[
S_{\text{ssdeep}}(C_i, C_j) \in [0,100].
\]
While Jaccard ignores token order, \texttt{ssdeep} is sensitive to preserved contiguous code regions and can identify repositories that share large implementation blocks despite localized modifications.

We treat the two metrics as complementary signals rather than combining them into a single score. High Jaccard similarity indicates strong lexical overlap, whereas high \texttt{ssdeep} similarity indicates preserved code structure. We use both scores to identify candidate clone pairs, but assign final clone labels through manual verification.

\begin{table}[t]
\centering
\caption{Summary statistics of the collected MCP and Skills datasets.}
\label{tab:stats}
\resizebox{\textwidth}{!}{%
\begin{tabular}{lrrr}
\toprule
Dataset & \# Repositories & \# Tools / Skills & Avg. \# per Repository \\
\midrule
MCP repositories with tools & 7,508 & 87,564 & 11.66 \\
Skills repositories & 1,353 & 12,447 & 9.20 \\
\bottomrule
\end{tabular}%
}
\end{table}

\begin{table}[t]
\centering
\caption{Distribution of repositories by primary programming language.}
\label{tab:languages}
\resizebox{0.3\textwidth}{!}{%
\begin{tabular}{lr}
\toprule
Language & \# Repositories \\
\midrule
Python & 2,990 \\
TypeScript & 2,208 \\
JavaScript & 1,219 \\
Go & 453 \\
Java & 225 \\
Rust & 133 \\
C\# & 114 \\
Other & 166 \\
\midrule
Total & 7,508 \\
\bottomrule
\end{tabular}%
}
\end{table}

\begin{table}[t]
\centering
\caption{Top-$k$ developer contribution shares in the MCP and Skills ecosystems.}
\begin{tabular}{llrrrr}
\toprule
Ecosystem & Top-$k$ & Tools & Share & Repos & Repo Share \\
\midrule
MCP    & 10 & 24,189 & 27.6\% & 194 & 2.6\% \\
MCP    & 20 & 27,687 & 31.6\% & 222 & 3.0\% \\
MCP    & 50 & 34,413 & 39.3\% & 310 & 4.1\% \\
\midrule
Skills & 10 & 3,627  & 29.1\% & 11 & 0.8\% \\
Skills & 20 & 4,335  & 34.8\% & 22 & 1.6\% \\
Skills & 50 & 5,667  & 45.5\% & 56 & 4.1\% \\
\bottomrule
\end{tabular}
\label{tab:developer-concentration}
\end{table}

\begin{figure}[t]
    \centering
    \begin{minipage}[b]{0.49\textwidth}
        \centering
        \includegraphics[width=\linewidth]{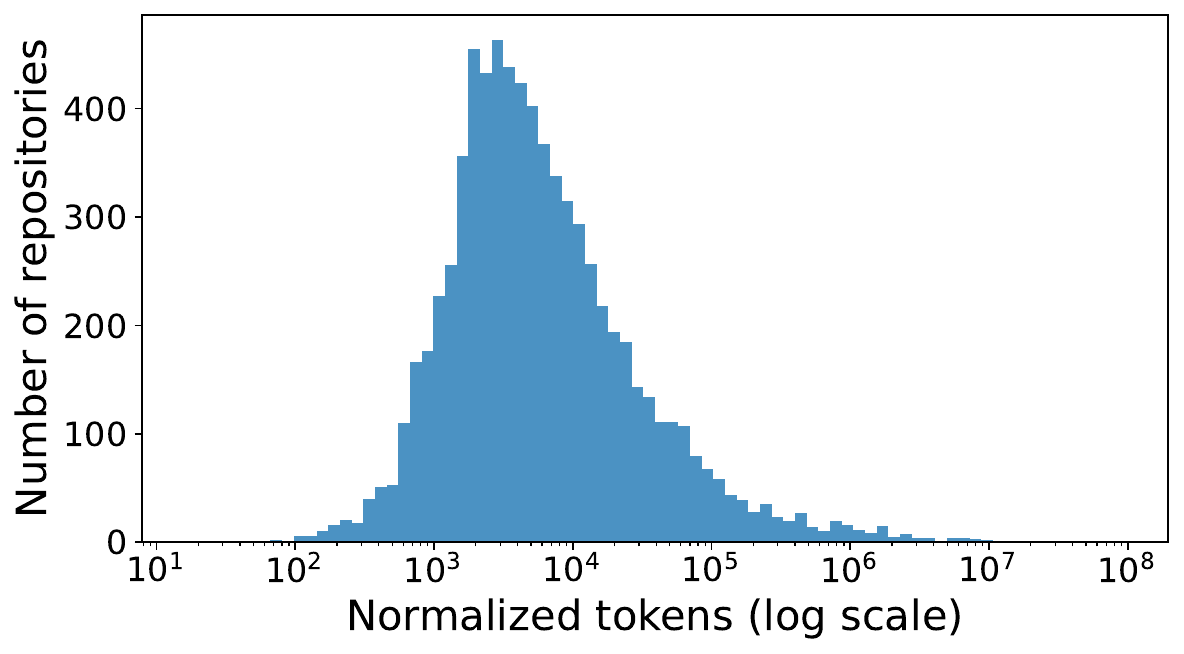}
        \centerline{(a) MCP repositories}
    \end{minipage}
    \hfill
    \begin{minipage}[b]{0.49\textwidth}
        \centering
        \includegraphics[width=\linewidth]{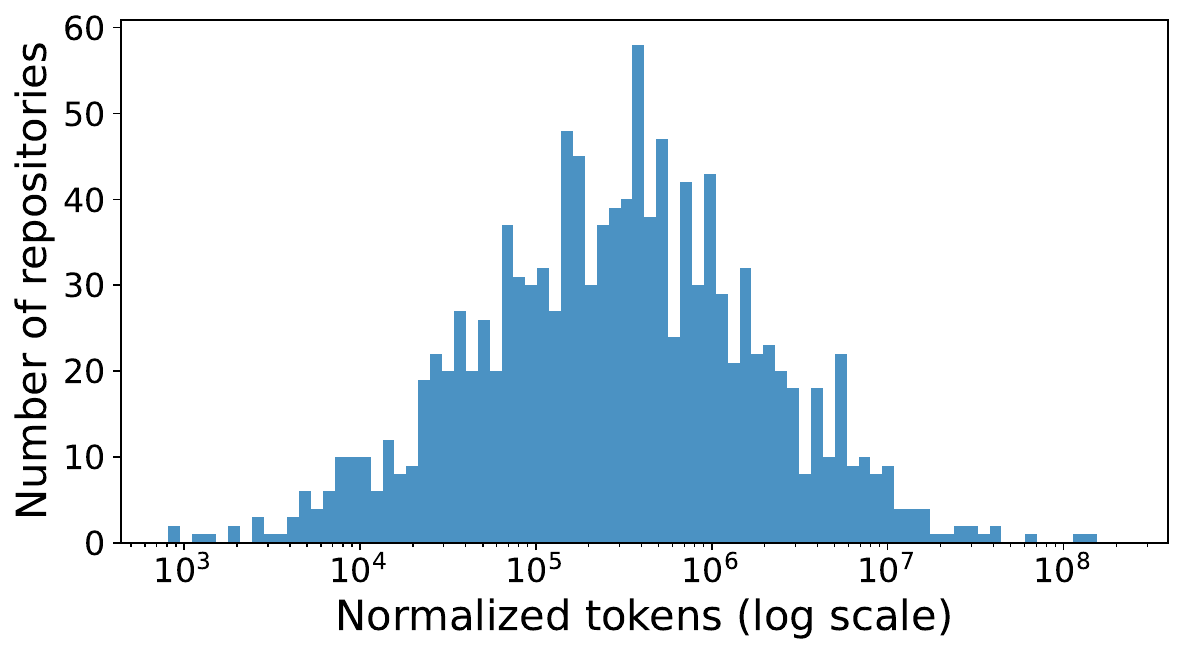}
        \centerline{(b) Skills repositories}
    \end{minipage}
    \caption{Distribution of MCP and Skills repository sizes measured by normalized source tokens.}
    \label{fig:code-size}
\end{figure}

\begin{figure}[t]
    \centering
    \includegraphics[width=1\textwidth]{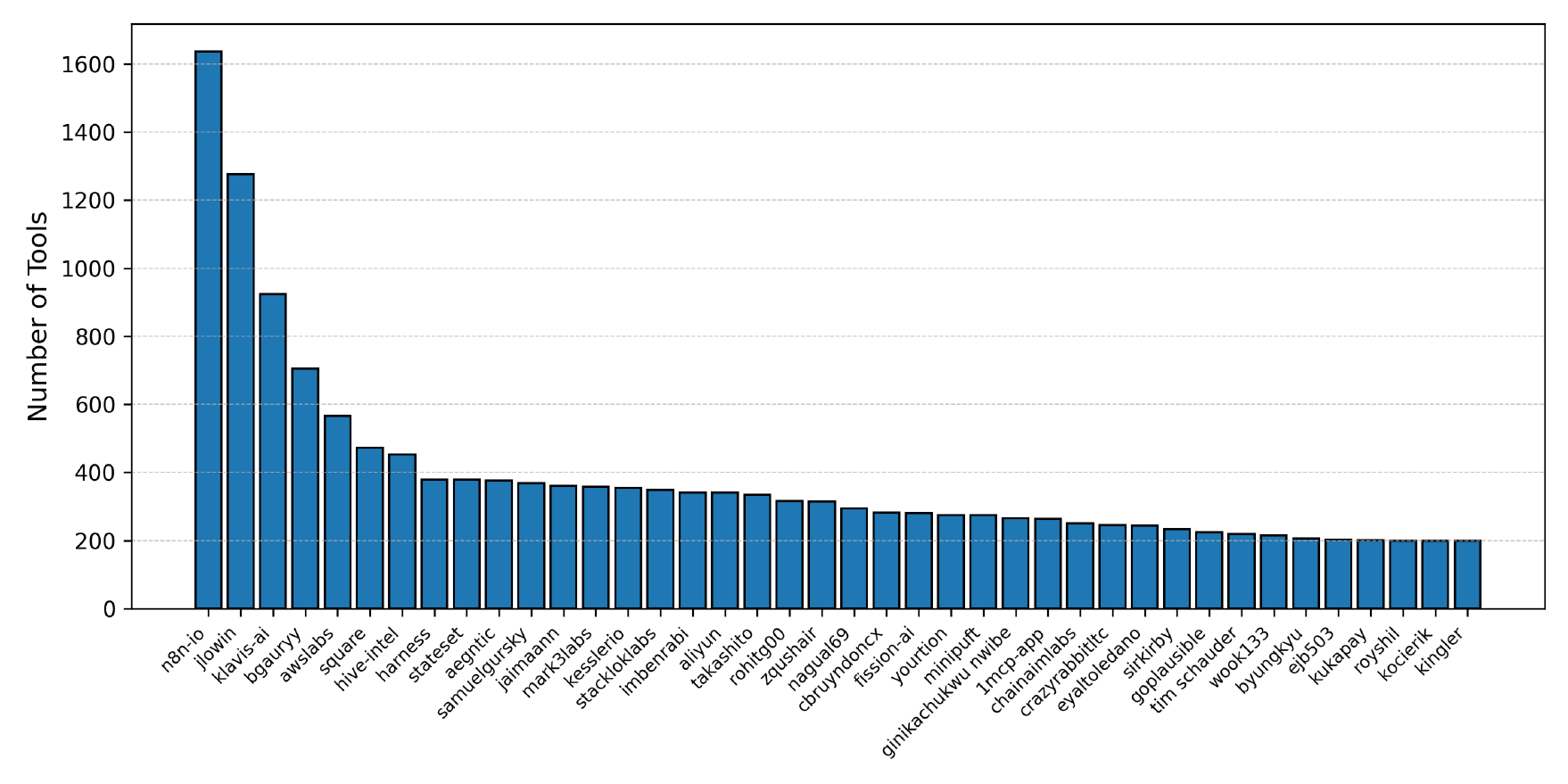}
    \caption{Distribution of tool counts for the top 40 authors in the MCP tool ecosystem.}
\end{figure}

\begin{figure}[t]
    \centering
    \includegraphics[width=1\textwidth]{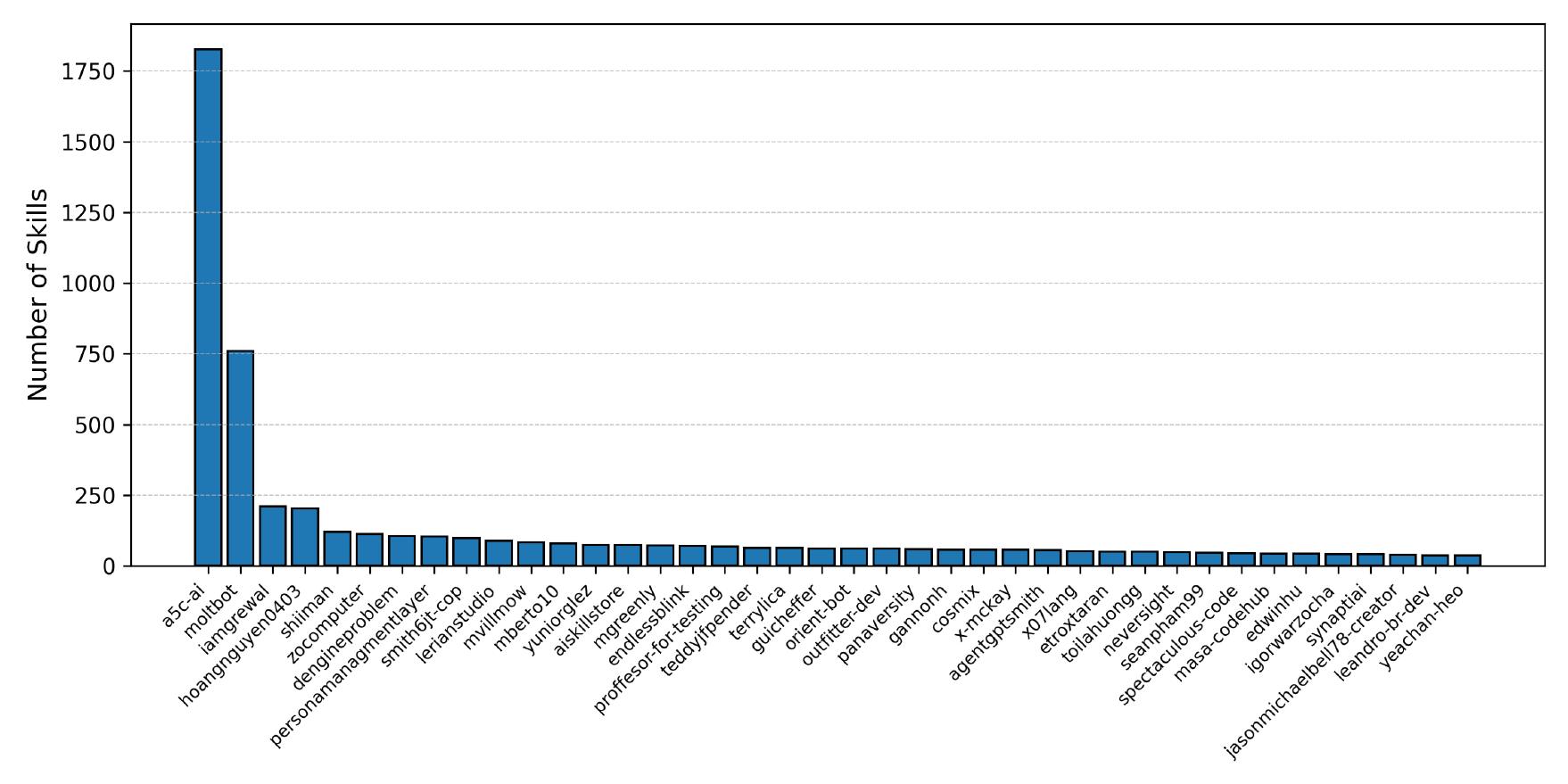}
    \caption{Distribution of skill counts for the top 40 authors in the Skills tool ecosystem.}
\end{figure}

\begin{figure}[t]
    \centering
    \begin{subfigure}[b]{0.49\textwidth}
        \centering
        \includegraphics[width=\linewidth]{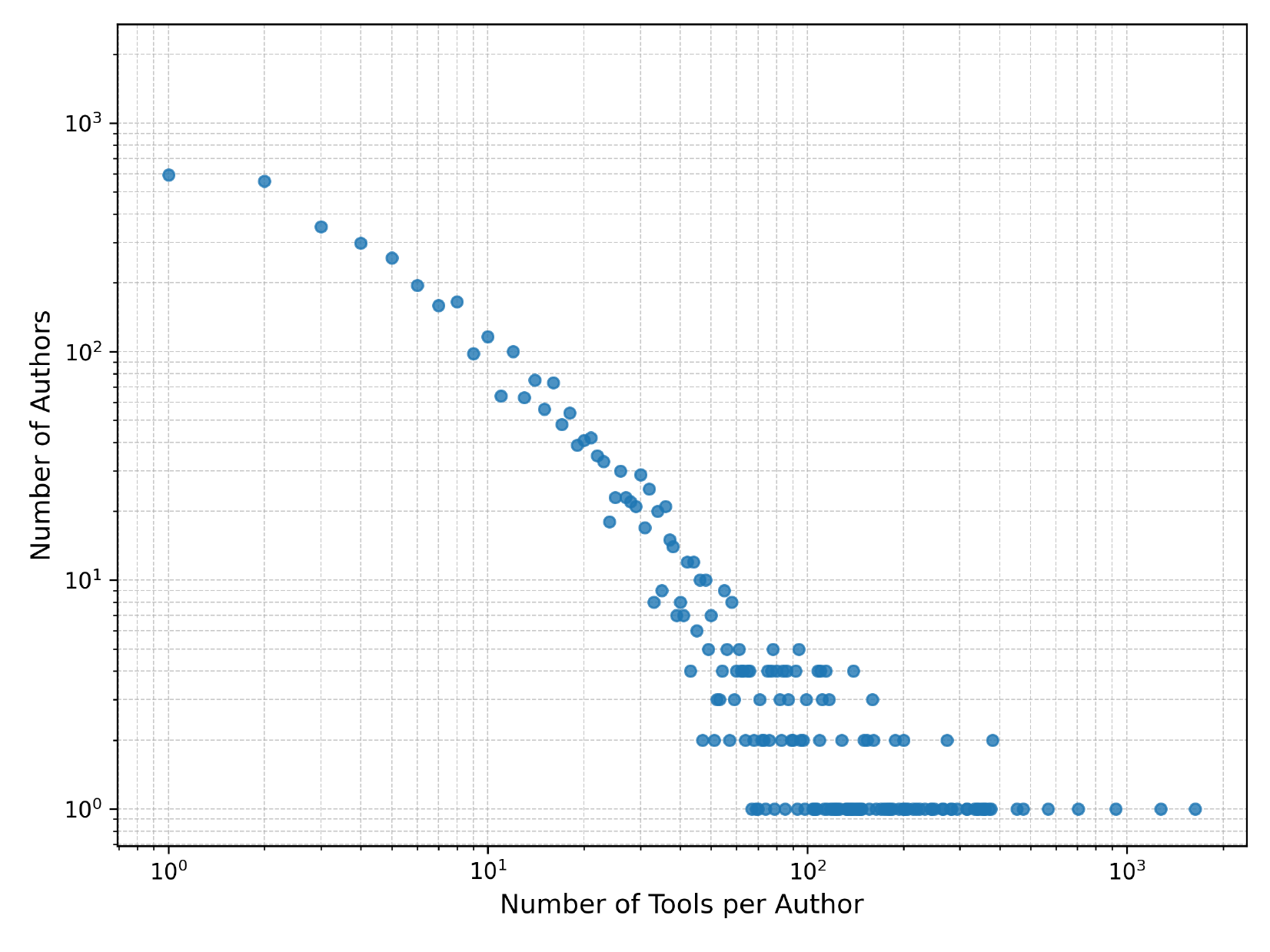}
        \phantomsubcaption
        \label{fig:mcp-author-frequency}
        \centerline{(a)}
    \end{subfigure}
    \hfill
    \begin{subfigure}[b]{0.49\textwidth}
        \centering
        \includegraphics[width=\linewidth]{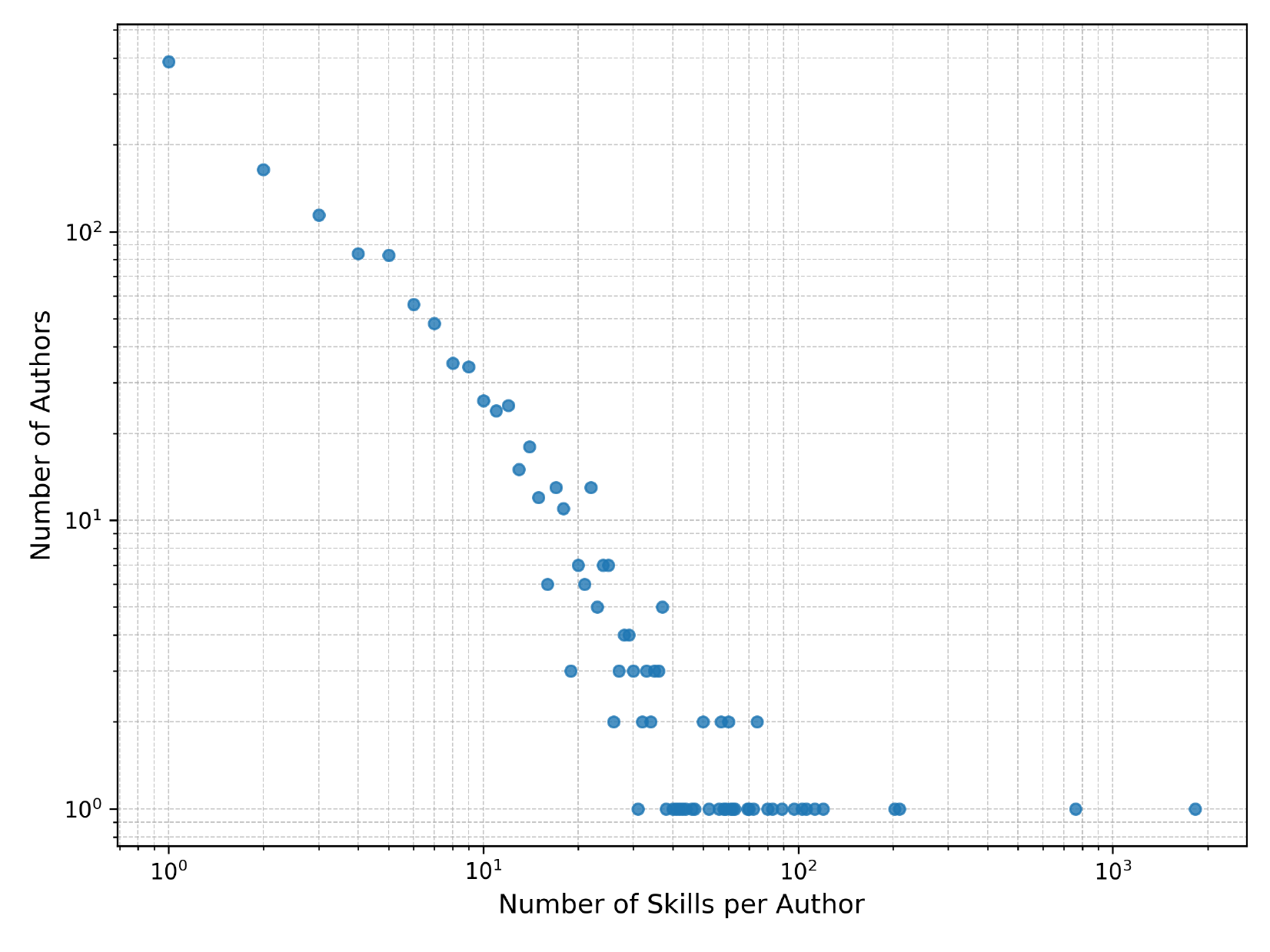}
        \phantomsubcaption
        \label{fig:skills-author-frequency}
        \centerline{(b)}
    \end{subfigure}
    \caption{
    Log-log distributions of developer contribution frequency.
    (a) MCP ecosystem.
    (b) Skills ecosystem.
    The x-axis represents total tool or skill contributions per developer, and the y-axis represents the number of developers with that contribution count.
    }
    \label{fig:author-frequency-distribution}
\end{figure}

\begin{figure}[t]
    \centering
    \includegraphics[width=1\textwidth]{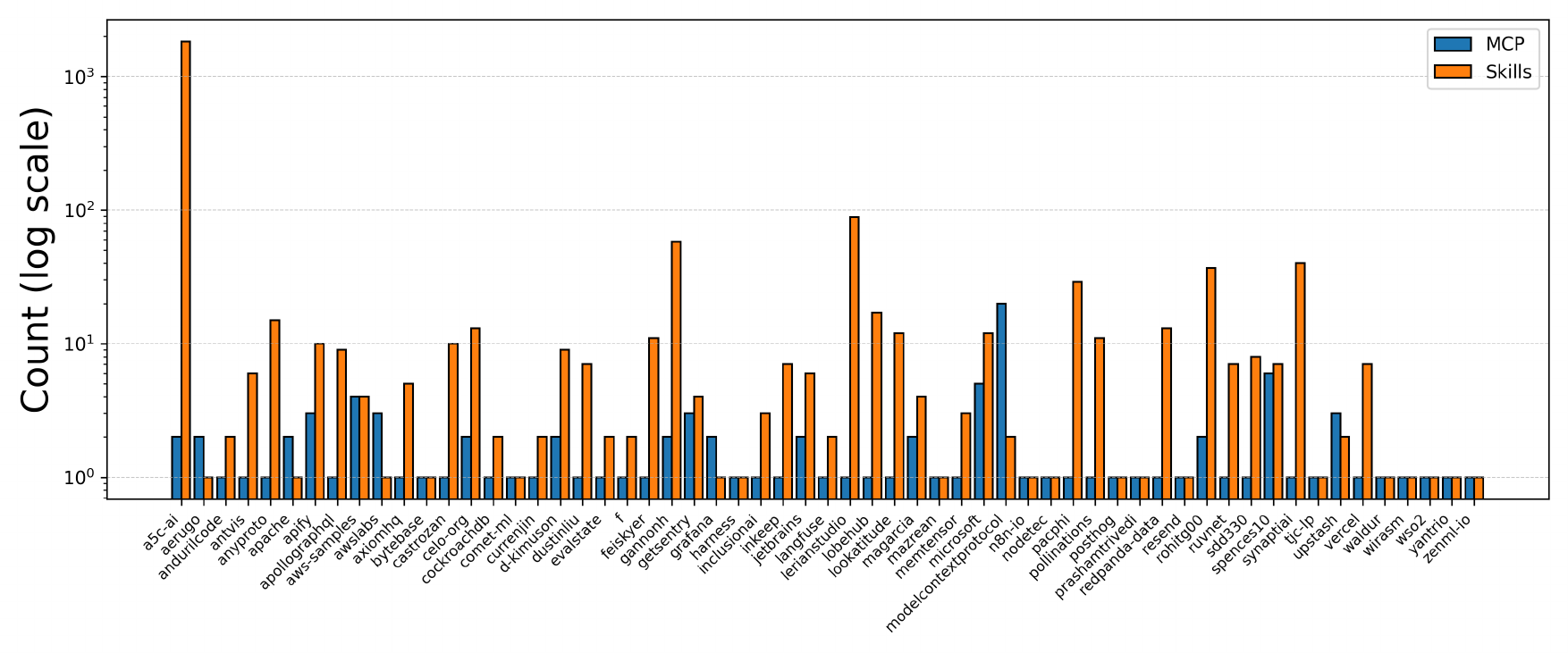}
    \caption{Distribution of MCP and Skills tools for authors present in both ecosystems (log scale).}
\end{figure}

\begin{figure}[t]
    \centering
    \begin{subfigure}[b]{0.4\textwidth}
        \centering
        \includegraphics[width=\linewidth]{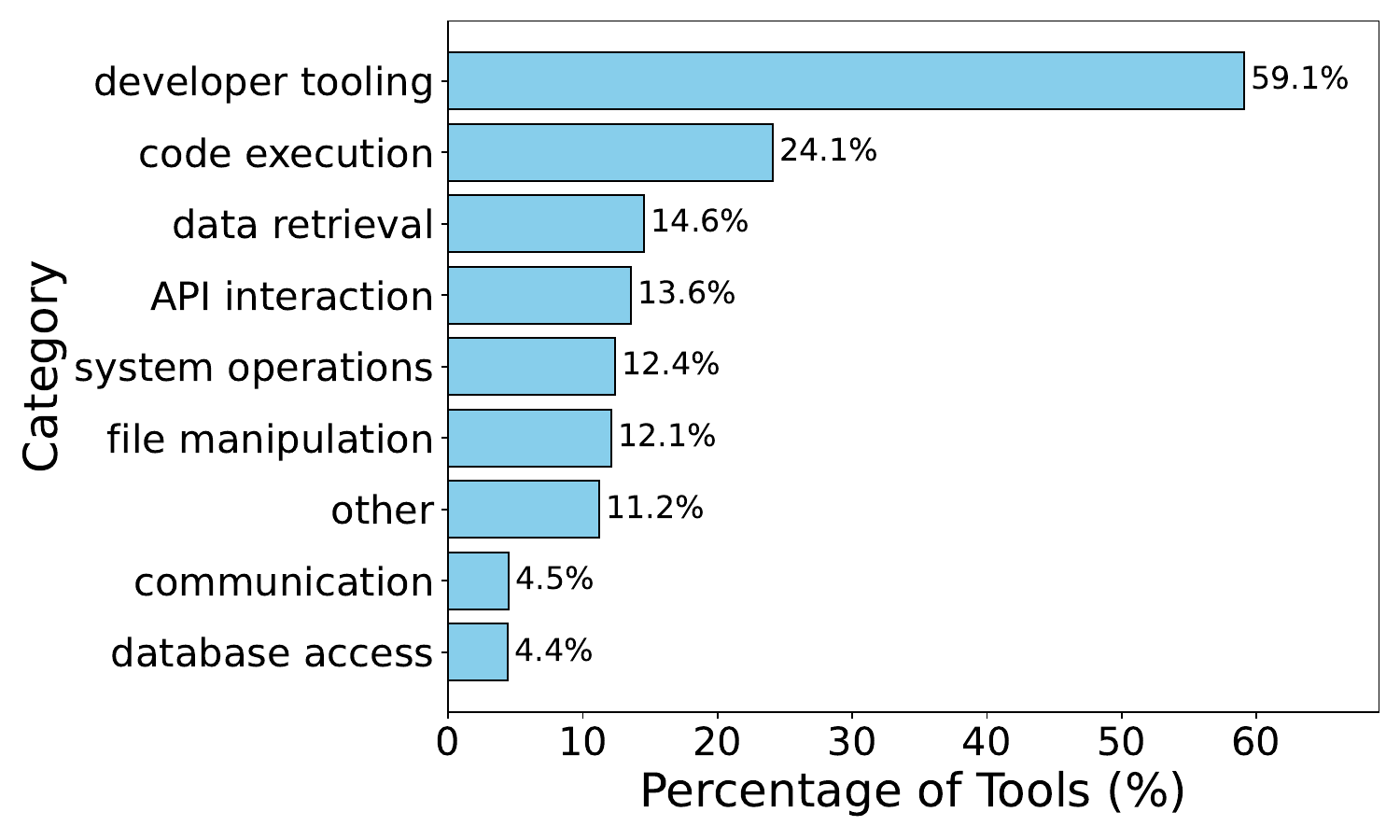}
        \caption{}
        \label{fig:skills-functionality}
    \end{subfigure}
    \begin{subfigure}[b]{0.4\textwidth}
        \centering
        \includegraphics[width=\linewidth]{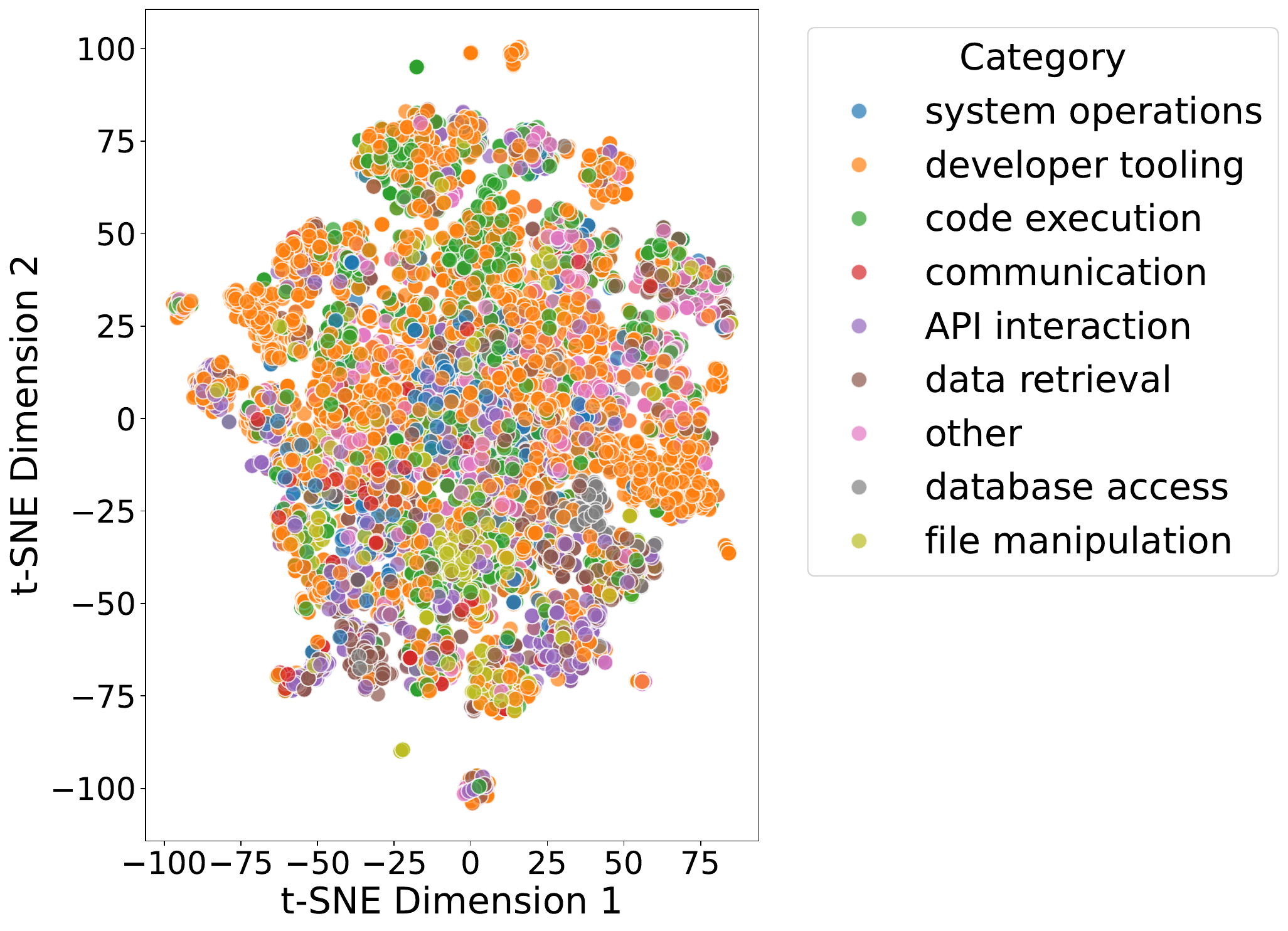}
        \caption{}
        \label{fig:skills-tsne}
    \end{subfigure}

    \caption{
    Functionality and description-space analysis of Skills.
    (a) Skills functionality distribution.
    (b) t-SNE projection of TF-IDF representations for Skills descriptions.
    Category percentages may sum to more than 100\% because skills can receive multiple labels.
    The t-SNE plot visualizes metadata-level semantic diversity only; it does not measure implementation diversity.
    }
    \label{fig:skills-functionality-metadata}
\end{figure}

\begin{table}[t]
\centering
\caption{
Prevalence of high-similarity candidate pairs across comparison groups.
Candidate pairs are repository pairs whose similarity scores exceed the calibrated threshold.
``Repos involved'' counts the unique repositories appearing in these pairs.
``Largest cluster'' is the size of the largest group of repositories connected through high-similarity pairwise relationships.
}
\label{tab:clone-candidate-prevalence}
\resizebox{0.7\textwidth}{!}{%
\begin{tabular}{llrrrr}
\toprule
Metric & Group & Threshold & Candidate pairs & Repos involved & Largest cluster \\
\midrule
Jaccard & MCP--MCP        & $\geq$ 80 & 758 & 441 & 38 \\
Jaccard & Skills--Skills  & $\geq$ 80 & 2   & 3   & 3  \\
Jaccard & MCP--Skills     & $\geq$ 80 & 4   & 8   & 2  \\
\midrule
\texttt{ssdeep} & MCP--MCP        & $\geq$ 80 & 517 & 391 & 25 \\
\texttt{ssdeep} & Skills--Skills  & $\geq$ 80 & 94  & 36  & 12 \\
\texttt{ssdeep} & MCP--Skills     & $\geq$ 80 & 37  & 50  & 7  \\
\bottomrule
\end{tabular}%
}
\end{table}

\subsection{Manual Verification Rubric}
\label{app:verification-rubric}

For each sampled repository pair, annotators apply the following rubric.

\begin{enumerate}
    \item \textbf{Repository structure.} Compare directory layout, file organization, and naming conventions. Highly similar structures are treated as evidence for possible reuse but are not sufficient for a clone label.

    \item \textbf{Core implementation overlap.} Inspect key source files excluding manifests, generated files, and trivial boilerplate. We look for shared function definitions, control flow, API usage, tool registration logic, request/response handling, authentication code, and module composition.

    \item \textbf{Boilerplate filtering.} Discount overlap explained only by standard framework scaffolding, dependency files, package metadata, generated files, or initialization code. A clone label requires similarity beyond generic setup code.

    \item \textbf{Functional equivalence.} Check whether the repositories expose equivalent tool behavior or implement the same core functionality, even if names, comments, or local code organization differ.

    \item \textbf{Divergence assessment.} Distinguish superficial changes from substantive rewrites. Renaming, configuration edits, minor parameter changes, or localized refactoring are compatible with a clone label; major architectural changes are not.

    \item \textbf{Final label.} Assign \texttt{clone} if multiple forms of evidence indicate substantive implementation reuse; otherwise assign \texttt{non-clone}.
\end{enumerate}

\subsection{Wilson Confidence Intervals}
\label{app:wilson-ci}

For each similarity bucket, let \(k\) denote the number of verified clones among \(n\) manually inspected pairs, and let \(\hat{p}=k/n\) be the observed clone proportion. We report a 95\% Wilson score confidence interval:
\[
\frac{
\hat{p} + \frac{z^2}{2n}
\pm
z \sqrt{
\frac{\hat{p}(1-\hat{p})}{n}
+
\frac{z^2}{4n^2}
}
}{
1 + \frac{z^2}{n}
},
\quad z=1.96.
\]
We use Wilson intervals rather than standard normal approximations because several buckets have small sample sizes or observed proportions near 0 or 1, where normal intervals can be unstable or degenerate.

\subsection{Limitations}
Our study focuses exclusively on public MCP and Skills repositories, thus our findings may not fully capture cloning dynamics within proprietary agentic ecosystems. Additionally, our analysis of MCP tool descriptions relies on public marketplace listings rather than direct repository extraction, which limits the scale of our metadata analysis. Furthermore, while our auditing pipeline leverages lexical and fuzzy-structural similarity metrics to effectively identify near-duplicates, it may overlook complex semantic clones where the underlying code has been heavily refactored. Future research should expand to other ecosystems, extract tool-level metadata directly from repositories, and incorporate advanced semantic clone detection techniques to provide a more comprehensive assessment of potential cloning.

\end{document}